\documentclass[onecolumn,amsmath,noshowkeys,noshowpacs,amssymb,nofootinbib]{revtex4}

\usepackage{graphicx,color}

\usepackage{amsbsy}
\usepackage{amsthm,mathrsfs,amsopn}
\usepackage{mathtools}
\usepackage{cases}

\newtheorem{remark}{Remark}

\begin{document}
%\title{Turing instability in coupled nonlinear relativistic heat equations}
\title{\textcolor{black}{Finite propagation enhances Turing patterns in reaction-diffusion networked systems}}
\author{Timoteo Carletti$^*$ \& Riccardo Muolo}

\affiliation{Department of Mathematics \& naXys, Namur Institute for Complex Systems, University of Namur, rue Graf\'e 2, 5000 Namur, Belgium}
\affiliation{$^*$timoteo.carletti@unamur.be}

\begin{abstract} 
We hereby develop the theory of Turing instability for reaction-diffusion systems defined on complex networks \textcolor{black}{assuming finite propagation}. Extending to networked systems the framework introduced by Cattaneo in the 40's, we remove the unphysical assumption of infinite propagation velocity holding for reaction-diffusion systems, thus allowing to propose a novel view on the fine tuning issue and on existing experiments. We analytically prove that Turing instability, stationary or wave-like, emerges for a much broader set of conditions, e.g., once the activator diffuses faster than the inhibitor or even in the case of inhibitor-inhibitor systems, overcoming thus the classical Turing framework. Analytical results are compared to direct simulations made on the FitzHugh-Nagumo model, extended to the relativistic reaction-diffusion framework with a complex network as substrate for the dynamics. 
\end{abstract}

%\pacs{89.75.Hc, 89.75.Kd,89.75.Fb}

%\keywords{Turing instability, Turing patterns, relativistic heat equation, hyperbolic reaction-diffusion systems, Spatio-temporal patterns, complex networks, Turing waves}
\maketitle

\vspace{0.8cm}

\section{Introduction}
\label{sec:intro}
\noindent
A blossoming of regular spatio-temporal patterns can be observed in Nature. These are the signature of self-organised processes where ordered structures emerge from disordered ones~\cite{Nicolis1977,Murray2001}. Very often, the interaction among the microscopic units, by which the system is made of, can be modelled by means of reaction-diffusion equations that \textcolor{black}{describe} the deterministic evolution of the concentrations both in time and space, the latter being a regular substrate~\cite{Murray2001} or a discrete one, e.g., a complex network~\cite{pastorsatorrasvespignani2010}. Spatially homogeneous equilibria of a reaction-diffusion system may undergo a symmetry breaking instability, when subjected to a heterogeneous perturbation, eventually driving the system towards a patchy, i.e., spatially heterogeneous, stationary or oscillatory solution, as firstly explained by Alan Turing~\cite{Turing} in the 50's \textcolor{black}{ and corroborated experimentally almost four decades later \cite{Castets,DeKepper,Tompkins}.}

%%%%%%%%%%%%%%% End of first page %%%%%%%%%%%%%%%%%%%%%

\maketitle

\noindent
Nowadays, applications of the Turing instability phenomenon go well beyond the original framework of the morphogenesis \textcolor{black}{or chemical reaction systems} and it stands for a pillar to explain self-organisation in Nature~\cite{Pecora_etal97,Strogatz01Nature,Pismen06}. The conditions for the emergence of Turing patterns have been elegantly grounded on the interplay between slow diffusing activators and fast diffusing inhibitors~\cite{GiererMeinhardt}; indeed this determines a local feedback, short range production of a given species, which should be, at the same time, inhibited at long ranges. Starting from these premises, scholars have been able to extend the original Turing mechanism to non-autonomous systems, e.g., evolving domains~\cite{CGM1999,PSGPBM2004} or time dependent diffusion and reaction rates~\cite{vangorder}, as well as discrete substrates, e.g., lattices~\cite{OthmerScriven74} or complex networks~\cite{NM2010}, and their generalisation, e.g., directed networks~\cite{Asllani2014NC}, multiplex networks~\cite{Asllani2014} and recently to time varying networks~\cite{PABFC2017, vangorder2}. {The interested reader can consult the recent review~\cite{KGMK2021} for a modern perspective on Turing instability.}

As previously observed, at the root of Turing instability there is a reaction-diffusion process which is thus grounded on a (nonlinear)``heat equation'', namely a parabolic partial differential equation (PDE)  {with a nonlinear source term}. The latter PDE is characterised by an infinite fast propagation of the initial datum along the supporting medium and thus it can accurately model the physical phenomenon only in cases of very large diffusivity, $D\gg 1$. To overcome this drawback, scholars have considered more realistic frameworks. In particular Cattaneo proposed in the 1948 to modify the constitutive equation (Fick's first law) by including a relaxation term with some given characteristic inertial time, $\tau>0$. Operating in this framework, Fick's second law returns a modified diffusion equation allowing also for a second derivative with respect to time~\cite{Cattaneo1948,Cattaneo1958,CL1996,MFH}. The resulting equation is nowadays known in the literature as the Cattaneo equation as well as the telegraph equation, the damped nonlinear Klein-Gordon equations or the relativistic heat equation, depending on the research field {and on the feature one is interested to emphasise}~\cite{MRB2014}. In any case its main characteristic is to exhibit a finite propagation velocity, $v=\sqrt{D/\tau}$, and moreover in the limit of arbitrarily small relaxation time, $\tau\rightarrow 0$, one recovers Fick's second law and thus a {parabolic} reaction-diffusion model with an infinite propagation velocity. \textcolor{black}{Our focus being the role of the finite propagation, we will hereby name such framework finite-velocity as done in~\cite{Masoliver} or sometimes {\em relativistic heat equation} although no Lorentz phenomena are at play, to recall the existence of a maximal allowed velocity as for the speed of light in relativity theory; we thus operated a different choice with respect to~\cite{ZH2016,CurroValenti}, where the name ``hyperbolic reaction-diffusion equations'' has been preferred.}

The aim of this paper is to study the conditions for the onset of Turing instability, being them stationary or oscillatory patterns, for a reaction-diffusion system defined on top of a {\em complex network} and modified according to the {\em Cattaneo recipe}, to allow for a finite propagation velocity (Section~\ref{sec:RRDnet}). We thus consider two different species {populating} a network composed by $n$ nodes. When species happen to share the same node, they interact via nonlinear functions $f(u_i,v_i)$ and $g(u_i,v_i)$, {being $u_i$ and $v_i$ the species concentrations at the $i$-th node}. On the other hand, they can diffuse across the available network links. The local currents, i.e., associated to each links, are assumed to satisfy a modified constitutive equation, Fick's first law, that includes a relaxation term with a given {\em inertial time}. Hence, the continuity equation, Fick's second law, allows to derive a modified local diffusion term, i.e., defined on the node. The latter, together with the reaction part, determine the hyperbolic reaction-diffusion system defined on top of a complex network, we will hereby be interested in.

Our work extends to the network case, the study presented in~\cite{ZH2016}, realised under the simplifying hypothesis of equal inertial times for the two species, namely $\tau_u=\tau_v$, and assuming a continuous substrate. Indeed, we hereby assume generic inertial times for each species, $\tau_u\neq\tau_v$. Let observe that our results established for a discrete substrate, can be straightforwardly extended to the continuous case and thus they complete the work done in~\cite{ZH2016} {to allow for different inertial times. This setting has been recently studied in~\cite{CurroValenti} in the framework of hyperbolic reaction-diffusion models with cross-diffusion defined on a continuous substrate. Differently from our approach of directly adapting the Cattaneo idea to the current flowing on each network link, the authors of~\cite{CurroValenti} have used Extended Thermodynamics~\cite{Muller2013}; the resulting characteristic polynomial (see below) we obtained is different from the one given in~\cite{CurroValenti} and it allows us to draw several interesting conclusions}. 

The Turing mechanism relies on the assumption of the existence of a stable homogeneous equilibrium that looses its stability once subjected to spatially heterogeneous perturbations, in presence of a diffusive term; for this reason, such process is also known as {\em diffusion-driven instability}. {The system can then exhibit stationary spatially heterogeneous solutions as well as time oscillatory ones.} The same mechanism can be proved to hold true in the new proposed framework of the hyperbolic reaction-diffusion systems defined on top of a complex network (Section~\ref{sec:TRelinst}). The dispersion relation, which ultimately signals the onset of the instability, is a function of the discrete spectrum of the Laplace matrix, namely the diffusion operator associated to the underlying network. \textcolor{black}{Let us mention that cross-diffusion is excluded.} The dispersion relation is obtained from the roots of the fourth order characteristic polynomial. To progress with the analytical understanding of the problem, we resort to the Routh-Hurwitz stability criterion~\cite{Routh1877,Hurwitz1895,Barnett1983}, allowing to prove the (in)stability feature of a real coefficients polynomial. Let us observe that this criterion is a widely used tool in dynamical systems and control theory (see, e.g.,~\cite{MFH}).

We have shown that the use of the inertial times strongly enlarges the \textcolor{black}{parameter} region for which Turing instability {and Turing-waves} can emerge, even beyond the classical Turing conditions of fast inhibitor and slow activator. For generic values of the inertial times, $\tau_u\neq \tau_v$, we have proved that Turing patterns can set up with a {\em fast activator} and {\em slow inhibitor}{, with species exhibiting the {\em same diffusion coefficients} and even with an {\em inhibitor-inhibitor} system}. As in these cases classical Turing instability cannot develop and being the latter solely due to the presence of the inertial times, we propose to call them {\em inertia-driven instability}, {that can result into stationary or wave-like phenomena}. Of course, the proposed framework allows to prove the existence of Turing instability also for an inhibitor diffusing faster than the activator, as for the non-relativistic framework.

In the particular case where both species have the same inertial time, we have shown that the stability of the homogeneous solution is conditional to the inertial time; indeed there exists a threshold, $\textcolor{black}{\tau_{\mathrm{max}}}$, beyond which the homogeneous equilibrium turns out to be unstable. The system exhibits thus patterns but they cannot be associated to Turing instability, even if they are indistinguishable from the latter. 
Moreover, the threshold $\textcolor{black}{\tau_{\mathrm{max}}}$ depends on the model parameters and there are combinations of the latter for which it is arbitrary large; stated differently, for such parameters the homogeneous equilibrium is always stable (with respect to the inertial time). 

The theoretical framework hereby proposed has been complemented with a dedicated numerical analysis of the FitzHugh-Nagumo model~\cite{fitzhugh, nagumo,Rinzel} (see Section~\ref{sec:FHN}), that is a nonlinear system often used as paradigm for the study of the emergence of Turing patterns~\cite{Ohta,Shoji,CarlettiNakao,siebert} as well as for synchronisation phenomena~\cite{Vragovic,AQIL20121615}. The FitzHugh-Nagumo model has thus been extended to the framework of hyperbolic reaction-diffusion networked systems. We have numerically found stationary patterns as well as synchronised oscillatory ones. We have also found new interesting solutions {for which the dispersion relation has limited predictive power; indeed we showed the existence of two sets of model parameters associated to similar dispersion relations, for which the unstable modes have a nontrivial complex component, but in one case the solution oscillates in time whereas in the second it converges to a stationary pattern.}

{The proposed framework tackles thus the issue of infinite propagation velocity for networked reaction-diffusion systems, and it is general enough to account for novel interesting results, strengthening the importance of self-organisation in nonlinear networked system. In particular the possibility to prove the emergence of Turing patterns, being the latter stationary or wave-like, in the case of activator diffusing faster than the inhibitor but also in the case of inhibitor-inhibitor systems, could provide new insights into the fine tuning problem~\cite{HG2021,KGMK2021} and propose a novel view on experimental results.}

\textcolor{black}{
\section{Reaction-diffusion system with finite propagation on networks}}
\label{sec:RRDnet}
\noindent
The aim of this section is to extend Cattaneo's idea to a discrete substrate, i.e. to deal with a networked system. We will briefly show how to modify networked reaction-diffusion systems in order to allow for a finite velocity of propagation.

Let us thus consider a network made of $n$ nodes and connected by a collection of $m$ undirected links allowing form pairwise exchanges among nodes. Such structure can be encoded into the $m\times n$ {\em incidence matrix}, $\mathbf{M}$. Let $e=(i,j)$ be the link connecting nodes $i$ and $j$, then $M_{e i}=1$, $M_{e j}=-1$ and $M_{e \ell}=0$ for all $\ell\neq i, j$. From this matrix we can build the {\em Laplace matrix}, $\mathbf{L}=-\mathbf{M}^{\top}\mathbf{M}$, where $^{\top}$ denotes the matrix transpose. The {\em Laplace matrix} is symmetric by construction and thus it admits a set of orthonormal eigenvectors, $\vec{\phi}^{(\alpha)}$, and real non-positive~\footnote{The matrix $\mathbf{L}$ is negative-semidefinite. Indeed let $\vec{\phi}^{(\alpha)}$ be any orthonormal eigenvector, then $(\vec{\phi}^{(\alpha)},\mathbf{L}\vec{\phi}^{(\alpha)})=\Lambda^{(\alpha)}$ and at the same time $(\vec{\phi}^{(\alpha)},\mathbf{L}\vec{\phi}^{(\alpha)})=-(\vec{\phi}^{(\alpha)},\mathbf{M}^{\top}\mathbf{M}\vec{\phi}^{(\alpha)})=-(\mathbf{M}\vec{\phi}^{(\alpha)},\mathbf{M}\vec{\phi}^{(\alpha)})=-||\mathbf{M}\vec{\phi}^{(\alpha)}||^2\leq 0$.} eigenvalues $\Lambda^{(\alpha)}$, for $\alpha=1,\dots,n$. By construction $\sum_j L_{ij}=0$, hence the largest eigenvalue is $\Lambda^{(1)}=0$ associated to the eigenvector $\vec{\phi}^{(1)}=(1,\dots,1)^{\top}/\sqrt{n}$. The diagonal element $-{L}_{ii}$ defines the nodes degree, say $k_i$, namely the number of incidents links of the $i$-th node; hence we can rewrite $\mathbf{L}=\mathbf{A}-\mathbf{D}$, where $\mathbf{D}=\mathrm{diag}(k_1,\dots,k_n)$ and $\mathbf{A}$ is the {\em adjacency matrix}, that is $A_{ij}=1$ if and only if nodes $i$ and $j$ are connected, encoding thus the coupling network.

Let us now focus on the diffusion of a single species in the network, the generalisation to more species being a direct extension. Let $\vec{u}(t)=(u_1(t),\dots,u_n(t))^{\top}$ to denote the state of the system at time $t$, where $u_i(t)$ is the density of the species in node $i$ at time $t$. Let $e=(i,j)$ be a link in the network and let $\chi_e(t)$ be the current flowing through it at time $t$; then, borrowing the constitutive equation, namely Fick's first law, from the continuous framework, we can state that 
\begin{equation}
\label{eq:Fick1net}
\chi_e(t)=-D_u\left[u_j(t)-u_i(t)\right]\equiv D_u\left[\mathbf{M}\vec{u}(t)\right]_e \, ,
\end{equation}
that is, the current is proportional to the difference of the densities in the nodes forming the link and flowing from higher concentrations to lower ones~\footnote{Let us observe that, despite the different sign in front of Eq.~\eqref{eq:Fick1net}, the latter is the analogous of the Fick's first law in the continuous setting: the current flows from regions of higher concentration to regions of lower one. Indeed, once we fix the link ``ordering'' as $e=(i,j)$, then the current $\chi_e$ will be positive, i.e., respecting the link ordering if $u_i>u_j$, while the current will be negative, i.e., opposite to the link order if $u_i<u_j$.}, being $D_u$ the diffusion coefficient of species $u$. By defining the currents vector $\vec{\chi}=(\chi_{e_1},\dots,\chi_{e_m})^{\top}$, the continuity equation can be written as
\begin{equation}
\label{eq:contnet}
\frac{du_i}{dt}(t)=-\left[\mathbf{M}^{\top} \vec{\chi}(t)\right]_i\, ,
\end{equation}
namely the variation of $u_i$ is proportional to the sum of the currents entering and exiting from node $i$. The classical Fick second law follows by combining the above equations:
\begin{equation}
\label{eq:fick2}
\frac{d\vec{u}}{dt}(t)=-\mathbf{M}^{\top} \vec{\chi}=-D_u\mathbf{M}^{\top}\mathbf{M}\vec{u}=D_u\mathbf{L}\vec{u}\, ,
\end{equation}
where we can realise~\cite{OthmerScriven74,NM2010} that $\mathbf{L}$ replaces the second order differential operator used in the continuous substrate case and thus the model given by~\eqref{eq:fick2} exhibits infinite propagation velocity.

To overcome this problem we modify, as Cattaneo did, the constitutive equation~\eqref{eq:Fick1net} by introducing a relaxation factor with some characteristic inertial time $\tau_u>0$, namely 
\begin{equation}
\label{eq:Fick1netrelax}
\chi_e(t)+\tau_u\frac{d\chi_e}{dt}(t)=D_u\left[\mathbf{M}\vec{u}(t)\right]_e\, .
\end{equation}
Combining this equation with the continuity equation~\eqref{eq:contnet} allows us to obtain
\begin{eqnarray}
\label{eq:fick2relax}
\frac{d\vec{u}}{dt}(t)&=&-\mathbf{M}^{\top} \vec{\chi}=-\mathbf{M}^{\top} \left[-\tau_u\frac{d\vec{\chi}}{dt}+D_u\mathbf{M}\vec{u}(t)\right]=\tau_u\frac{d\mathbf{M}^{\top}\vec{\chi}}{dt}-D_u\mathbf{M}^{\top}\mathbf{M}\vec{u}(t)\notag\\
&=&-\tau_u\frac{d^2\vec{u}}{dt^2}+D_u\mathbf{L}\vec{u}(t)\, ,
\end{eqnarray}
eventually providing the generalised Cattaneo equation defined on networks
\begin{equation}
\label{eq:cattnet}
\frac{d\vec{u}}{dt}(t)+\tau_u\frac{d^2\vec{u}}{dt^2}(t)=D_u\mathbf{L}\vec{u}(t)\, .
\end{equation}
{The latter can be seen as a modification of the ``heat equation'' defined on  network by the inclusion of a second order time derivative, returning thus a relativistic or hyperbolic heat equation.}

Consider now two different species populating a network composed by $n$ nodes and let us denote by $u_i$ and $v_i$, $i=1,\dots,n$, their respective concentrations on node $i$. When species happen to share the same node, they interact via nonlinear functions $f(u_i,v_i)$ and $g(u_i,v_i)$. On the other hand, they can diffuse across the available network links accordingly to the modified Cattaneo equation~\eqref{eq:cattnet}. The model can hence be mathematically cast in the form
\begin{equation}
\label{eq:TRelnet}
\begin{dcases}
\frac{du_i}{dt}+\tau_u \frac{d^2u_i}{dt^2}&=f(u_i,v_i)+D_u\sum_{j=1}^{n}L_{ij} u_j  \\ 
\frac{dv_i}{dt}+\tau_v \frac{d^2v_i}{dt^2}&=g(u_i,v_i)+D_v\sum_{j=1}^{n}L_{ij} v_j 
\end{dcases} \quad\forall i=1,\dots,n\, ,
\end{equation}
where $D_u>0$ (resp. $D_v>0$) is the diffusion coefficients of species $u$ (reps. $v$) and $\tau_u>0$ (resp. $\tau_v>0$) the inertial time for species $u$ (resp. $v$).

\textcolor{black}{
\section{Turing instability in networked reaction-diffusion systems with finite propagation}}
\label{sec:TRelinst}
\noindent
The Turing mechanism is the result of a diffusion-driven instability, namely an homogeneous stable equilibrium of the reaction-diffusion system turns out to be unstable, with respect to inhomogeneous spatial perturbations, once the diffusion is at play. The aim of this section is to determine the conditions for such instability to develop in the relativistic\textcolor{black}{, i.e., in presence of a maximal allowed velocity,} reaction-diffusion systems defined on networks given by Eq.~\eqref{eq:TRelnet}.

Let us hence assume there exists an homogeneous solution of~\eqref{eq:TRelnet}, that is $u_i(t)=u_0$ and $v_i(t)=v_0$ for all $i=1,\dots,n$ and $t>0$. Namely $u_0$ and $v_0$ should satisfy $f(u_0,v_0)=g(u_0,v_0)=0$. {Being the latter equilibrium solely determined by the reaction terms, it happens to be also an equilibrium for the non-relativistic system.} Let us denote by $\delta u_i(t)=u_i(t)-u_0$ and $\delta v_i(t)=v_i(t)-v_0$ the perturbations from the homogeneous solution. In order to determine the time evolution of the latter, we use~\eqref{eq:TRelnet}, keeping only the first order terms in the perturbation (the latter assumed to be small). We thus obtain
\begin{equation}
\label{eq:Cattaneo3}
\begin{dcases}
\frac{d\delta u_i}{dt}+\tau_u \frac{d^2\delta u_i}{dt^2}&=\partial_u f \delta u_i +\partial_v f \delta v_i + D_u\sum_{j=1}^{n}L_{ij} \delta u_j  \\ 
\frac{d\delta v_i}{dt}+\tau_v \frac{d^2\delta v_i}{dt^2}&=\partial_u g \delta u_i +\partial_v g \delta v_i +D_v\sum_{j=1}^{n}L_{ij} \delta v_j 
\end{dcases} \quad\forall i=1,\dots,n\, ,
\end{equation}
where we employed the fact that $\sum_j L_{ij}=0$ to {nullify the terms $\sum_jL_{ij}u_0$ and $\sum_jL_{ij}v_0$}. Let us also stress that throughout the rest of the section the partial derivatives, i.e., $\partial_u f\equiv \partial f/\partial u$ and similarly for the other ones, are evaluated at the homogeneous equilibrium $(u_0,v_0)$.

To progress with the analytical understanding, we develop the perturbations on the eigenbasis of the Laplace matrix $\delta u_i(t)=\sum_\alpha \hat{u}_\alpha(t) \phi^{(\alpha)}_i$ and $\delta v_i(t)=\sum_\alpha \hat{v}_\alpha (t)\phi^{(\alpha)}_i$. Inserting the latter into Eq.~\eqref{eq:Cattaneo3} we obtain the equation \textcolor{black}{describing} the evolution of the modes $\hat{u}_\alpha(t)$ and $\hat{v}_\alpha(t)$
\begin{equation}
\label{eq:Cattaneo4}
\begin{dcases}
\frac{d\hat{u}_\alpha}{dt}(t) + \tau_u \frac{d^2\hat{u}_\alpha}{dt^2}(t)&= \partial_uf\hat{u}_\alpha(t)+ \partial_vf\hat{v}_\alpha(t)+D_u\Lambda^{(\alpha)}\hat{u}_\alpha(t)\notag \\
\frac{d\hat{v}_\alpha}{dt}(t) + \tau_v \frac{d^2\hat{v}_\alpha}{dt^2}(t)&= \partial_ug\hat{u}_\alpha(t)+ \partial_vg\hat{v}_\alpha(t)+D_v\Lambda^{(\alpha)}\hat{v}_\alpha(t)
\end{dcases}\quad\forall i=1,\dots,n\, ,
\end{equation}
namely we end up with $n$ linear $2\times 2$ systems instead of the initial $2n\times 2n$ one. We further hypothesise 
%$\hat{u}_\alpha(t) \sim e^{\lambda_\alpha t}\hat{u}_\alpha$ and $\hat{v}_\alpha(t) \sim e^{\lambda_\alpha t}\hat{v}_\alpha$, and we eventually obtain
%\begin{eqnarray}
%\label{eq:Cattaneo5}
%\lambda_\alpha{\hat{u}_\alpha} + \tau_u \lambda_\alpha^2{\hat{u}_\alpha}&=& \partial_uf\hat{u}_\alpha+ \partial_vf\hat{v}_\alpha+D_u\Lambda^{(\alpha)}\hat{u}_\alpha\notag \\
%\lambda_\alpha{\hat{v}_\alpha} + \tau_v \lambda_\alpha^2{\hat{v}_\alpha}&= &\partial_ug\hat{u}_\alpha+ \partial_vg\hat{v}_\alpha+D_v\Lambda^{(alpha)}\hat{v}_\alpha\, ,
%\end{eqnarray}
%
%and to ensure the existence of a non trivial solution, the linear growth rate $\lambda_\alpha$ should solve
$\hat{u}_\alpha(t) \sim e^{\lambda_\alpha t}$ and $\hat{v}_\alpha(t) \sim e^{\lambda_\alpha t}$, and to ensure the existence of a nontrivial solution we eventually obtain that the linear growth rate $\lambda_\alpha$ should solve
\begin{equation}
\label{eq:Cattaneo6}
\det \left(
\begin{matrix}
\lambda_\alpha + \tau_u\lambda_\alpha^2-\partial_u f-\Lambda^{(\alpha)} D_u & -\partial_v f\\
-\partial_u g & \lambda_\alpha + \tau_v\lambda_\alpha^2-\partial_v g-\Lambda^{(\alpha)} D_v
\end{matrix}
\right)=0 \Leftrightarrow p_\alpha(\lambda_{\alpha})=0\, ,
\end{equation}
where the fourth degree {characteristic} polynomial is defined by
\begin{equation}
\label{eq:Cattaneo7}
p_\alpha(\lambda)=a\lambda^4+b\lambda^3+c_\alpha\lambda^2+d_\alpha\lambda+e_\alpha\, ,
\end{equation}
whose coefficients are given by
\begin{eqnarray}
a&=&\tau_u \tau_v \quad , \quad b=(\tau_u+\tau_v)\label{eq:AB}\\ c_\alpha &=& 1-\tau_u\partial_v g-\tau_v\partial_uf -\Lambda^{(\alpha)}\left(\tau_uD_v+\tau_v D_u\right)\label{eq:C}\\ d_\alpha &=& -\mathrm{tr}(J_0)-\Lambda^{(\alpha)}(D_v+D_u)\label{eq:D}\\ e_\alpha &=& \det(J_0)+\left(D_v\partial_u f+D_u\partial_v g\right)\Lambda^{(\alpha)}+D_uD_v\left(\Lambda^{(\alpha)}\right)^2 \label{eq:E}\, ,
\end{eqnarray}
being $J_0=\left(\begin{smallmatrix}\partial_u f & \partial_v f\\ \partial_u g & \partial_v g\end{smallmatrix}\right)$ the Jacobian of the \textcolor{black}{ sole reaction system without diffusive coupling, hence also named aspatial system}, evaluated at the homogeneous equilibrium $(u_0,v_0)$, $\mathrm{tr}(J_0)=\partial_u f+\partial_v g$ its trace and $\det(J_0)=\partial_u f\partial_v g-\partial_v f\partial_u g$ its determinant. The coefficients $a$ and $b$ are positive and do not depend on the index $\alpha$.

Turing instability arises if the homogeneous equilibrium $(u_0,v_0)$ is stable, namely if the four roots of the polynomial $p_1(\lambda)$ all have negative real part~\footnote{Inspecting Eq.~\eqref{eq:Cattaneo4} it is clear that the $\Lambda^{(1)}=0$ eigenvalue represents the behaviour of the aspatial system.}, while there exists at least one $\alpha >1$ for which the polynomial $p_\alpha(\lambda)$ does admit at least one root with positive real part. The root with the largest real part seen as a function of $\Lambda^{(\alpha)}$, is called in the literature the {\em dispersion relation}, $\lambda_\alpha:=\max_{i=1,\dots,4}\Re \lambda_i(\Lambda^{(\alpha)})$. Turing instability is thus equivalent to require $\lambda_1<0$ and $\lambda_{\alpha}>0$ for some $\alpha>1$, \textcolor{black}{hereby called critical roots}. {Indeed, because of the ansatz $\hat{u}_\alpha(t) \sim e^{\lambda_\alpha t}$ and $\hat{v}_\alpha(t) \sim e^{\lambda_\alpha t}$, the former implies an initial exponential divergence from the homogeneous equilibrium. \textcolor{black}{In the following we will also use information about the imaginary part of $\lambda_\alpha$, we thus define $\rho_\alpha:=\max_{i=1,\dots,4}\{\Im \lambda_i(\Lambda^{(\alpha)}): \lambda_\alpha>0\}$, i.e., the largest imaginary part of the critical roots.} In the case the imaginary part of the critical root is non-zero, $\rho_\alpha\neq 0$, we are in presence of a Turing-wave instability, and the perturbation initially exhibits a combination of exponential growth and oscillating behaviour. Eventually the nonlinearities of the model determine the final pattern, that could result to be stationary or wave-like one.}

To prove the existence of Turing instability for the system~\eqref{eq:TRelnet} we shall rely on the Routh-Hurwitz criterion~\cite{Routh1877,Hurwitz1895,Barnett1983}, providing necessary and sufficient conditions to prove that $p_1$ is stable~\footnote{Let us recall that, borrowed from the theory of the linear stability of dynamical systems, a polynomial is \textcolor{black}{(asymptotically)} stable if and only if all its roots have negative real part, while a polynomial is said to be unstable if there exists at least one root with positive real part.} while $p_{{\alpha}}$ is unstable for some ${{\alpha}}>1$

\vspace{1Em}
\begin{remark}[Connection with the relativistic reaction-diffusion system defined on a continuous substrate]
%\textbf{Remark} ({\em Connection with the relativistic reaction-diffusion system defined on a continuous substrate})
 As already remarked, the Laplace matrix $\mathbf{L}$ in Eq.~\eqref{eq:TRelnet} takes the place of the second order differential operator $\Delta=\sum_i\partial^2_{x_i}$. After linearising the resulting PDE system about the homogeneous equilibrium, the use of the periodic boundary conditions and the Fourier series, is equivalent to project the linear system onto the eigenfunctions of $\Delta$, that is $e^{i k x}$ (for $k\in\mathbb{Z}$, in the case of a $1$ dimensional spatial domain), whose eigenvalues are $-k^2$. Proceeding in this way, one can determine a polynomial similar to the one given in~\eqref{eq:Cattaneo7}, where we have to replace $\Lambda^{(\alpha)}$ by $-k^2$. {However, let us observe that now the spectrum of $\mathbf{L}$ is discrete and thus the dispersion relation for the networked system will be ``sampled'' from the one holding in the continuous case (see red dots on the blue curves in the following figures representing the dispersion relations). This may introduce finite size effects, as the continuous support case is capable to exhibit Turing patterns, while the networked one cannot because the Laplace spectrum has a gap that exactly avoids the region of positive dispersion relation. To control for this phenomenon, one should be able to relate topological features of the network to the Laplace spectrum~\cite{DMMS2018,MJPMG2019}.}
 \end{remark}
 %}

\subsection{Conditions for the stability of $p_1$}
\label{ssec:Condp0}
\noindent
The aim of this section is to introduce the conditions for the linear stability of the homogeneous solution of~\eqref{eq:TRelnet}.
As already noticed, the coefficients $a=\tau_u\tau_v$ and $b=\tau_u+\tau_v$ are positive, hence the necessary and sufficient conditions (see Appendix~\ref{sec:AppRHcrit}) to ensure the stability of $p_1$ are given by:
\begin{eqnarray}
1-\tau_u \partial_v g-\tau_v\partial_u f&>&0\label{cond11}\\
\mathrm{tr}(J_0)=\partial_uf+\partial_v g&<&0\label{cond12}\\
\det(J_0)=\partial_uf \partial_v g-\partial_vf \partial_u g&>&0\label{cond13}\\
(\tau_u+\tau_v)(1-\tau_u \partial_v g-\tau_v\partial_u f)+\tau_u\tau_v\mathrm{tr}(J_0) &>&0\label{cond14}\\
-\mathrm{tr}(J_0)\left[ (\tau_u+\tau_v)(1-\tau_u \partial_v g-\tau_v\partial_u f)+\tau_u\tau_v\mathrm{tr}(J_0)\right] -(\tau_u+\tau_v)^2\det(J_0)&>&0\label{cond15}\, ,
\end{eqnarray} 

Before proceeding {with the analysis in the general setting}, let us consider a special but relevant case, namely $\tau_u=\tau_v=\tau$. Assuming Eq.~\eqref{cond12} to hold true, then Eqs.~\eqref{cond11} and~\eqref{cond14} easily follow. Moreover, if $~4\det(J_0)<(\mathrm{tr}(J_0))^2$, then Eq.~\eqref{cond15} is always satisfied, while if $~4\det(J_0)>(\mathrm{tr}(J_0))^2$, the following upper bound for $\tau$ is obtained to satisfy~\eqref{cond15}: 
\begin{equation}
\label{eq:cond14taubound}
\tau < \textcolor{black}{\tau_{\mathrm{max}}}=\frac{-2\mathrm{tr}(J_0)}{\left[4\det(J_0)-(\mathrm{tr}(J_0))^2\right]}\, .
\end{equation}
This last results will be important in the following, because it states that the stability of the homogeneous equilibrium depends on $\tau$ (see panel b) in Fig.~\ref{fig:mualphaFHN}). More importantly, if $\tau$ is large enough, the system~\eqref{eq:TRelnet} exhibits patterns; they are {\em not emerging from a Turing mechanism} but {instead} from the {\em instability of the homogeneous equilibrium}{, observe that one cannot discriminate them with respect to Turing patterns by simple visual inspection.}

\subsection{Conditions for the instability of $p_\alpha$}
\label{ssec:Condpalpha}
\noindent
Using again the Routh-Hurwitz criterion we can prove the existence of (at least) \textcolor{black}{an} ${\alpha}>1$ for which $p_{{\alpha}}$ is unstable conditioned on the stability of $p_1$.
 
 Observe again that the coefficients $a$ and $b$ are positive. Moreover, by assuming Eqs.~\eqref{cond11} and~\eqref{cond12} to hold true and by recalling that $-\Lambda^{(\alpha)}>0$ for all $\alpha>1$, then $c_\alpha>0$ and $d_\alpha>0$ (see Eqs.~\eqref{eq:C} and~\eqref{eq:D}). In conclusion the unique coefficient of $p_\alpha$ that can be negative is $e_\alpha$. Hence (see Appendix~\ref{sec:AppRHcrit}) the instability can arise if one of the following couples of conditions is verified: 
%\begin{eqnarray}
%B&:=&-(\tau_u+\tau_v)(D_u+D_v)(1-\tau_u\partial_v g-\tau_v\partial_u f)+(\tau_u+\tau_v)\mathrm{tr}(J_0)(\tau_u D_v+\tau_v D_u)\notag\\&-&2\mathrm{tr}(J_0)(D_u+D_v)\tau_u\tau_v-(\tau_u+\tau_v)^2(D_v\partial_u f+D_u\partial_v g)>0\label{eq:inst21}\\
%B^2&-&4(D_v\tau_u-D_u\tau_v)^2\Big[-\mathrm{tr}(J_0)(\tau_u+\tau_v)(1-\tau_u\partial_vg-\tau_v\partial_uf)\notag\\&-&\tau_u\tau_v(\mathrm{tr}(J_0))^2-(\tau_u+\tau_v)^2\det(J_0)\Big]>0\label{eq:inst22}\, ,
%\end{eqnarray} 
\begin{subnumcases}{}
B:=-(\tau_u+\tau_v)(D_u+D_v)(1-\tau_u\partial_v g-\tau_v\partial_u f)+(\tau_u+\tau_v)\mathrm{tr}(J_0)(\tau_u D_v+\tau_v D_u)\notag\\\hspace{15em}-2\mathrm{tr}(J_0)(D_u+D_v)\tau_u\tau_v-(\tau_u+\tau_v)^2(D_v\partial_u f+D_u\partial_v g)>0\label{eq:inst21}\\
B^2-4(D_v\tau_u-D_u\tau_v)^2\Big[-\mathrm{tr}(J_0)(\tau_u+\tau_v)(1-\tau_u\partial_vg-\tau_v\partial_uf)-\tau_u\tau_v(\mathrm{tr}(J_0))^2-(\tau_u+\tau_v)^2\det(J_0)\Big]>0\label{eq:inst22}\, ,
\end{subnumcases} 
or
%\begin{eqnarray}
%D_u\partial_v g+D_v\partial_u f&>&0\label{eq:inst11}\\
%(D_u\partial_v g+D_v\partial_u f)^2-4{D_uD_v} \det(J_0)&>&0\label{eq:inst12}\, ,
%\end{eqnarray} 
\begin{subnumcases}{}
D_u\partial_v g+D_v\partial_u f>0\label{eq:inst11}\\
(D_u\partial_v g+D_v\partial_u f)^2-4{D_uD_v} \det(J_0)>0\label{eq:inst12}\, ,
\end{subnumcases} 

Let us observe that Eqs.~\eqref{eq:inst11} and~\eqref{eq:inst12} do not depend on $\tau_u$ and $\tau_v$ and are indeed the same conditions one imposes to obtain the Turing instability {in the classical, i.e., non-relativistic setting}~\cite{NM2010}. In particular, they require $D_v>D_u$. However, Eqs.~\eqref{eq:inst21} and~\eqref{eq:inst22} do not ask for such condition on the diffusivities, implying that the hypothesis of a finite propagation velocity allows to enlarge the \textcolor{black}{parameter} region for which Turing instability arises, {in particular allowing for $D_v\leq D_u$}. 

Based on the above one can conclude that if the patterns with positive inertial times are due to Eqs.~\eqref{eq:inst11} and~\eqref{eq:inst12}, then they persist also in the non-relativistic limit, $\tau_u\rightarrow 0$ and $\tau_v\rightarrow 0$. On the other hand, if the instability has been initiated by conditions Eqs.~\eqref{eq:inst21} and~\eqref{eq:inst22}, we can show (see Appendix~\ref{sec:Appnonrellim}) that in the non-relativistic limit, the patterns fade out and disappear for positive and sufficiently small inertial times.

To start our analysis, let us thus consider the case $D_u=D_v=D$. As already observed Eq.~\eqref{eq:inst11} cannot be satisfied having by Eq.~\eqref{cond12} the fact that $\mathrm{tr}(J_0)<0$, thus this cannot be a path towards Turing instability. On the other hand, let us reorganise terms and rewrite condition~\eqref{eq:inst21} as follows
\begin{equation*}
B|_{D_u=D_v=D}=-2D\left[ \tau_u+\tau_v +(\tau_u^2-\tau_v^2) \partial_uf +\tau_u(\tau_v-\tau_u) \mathrm{tr}(J_0)\right]\, ,
\end{equation*} 
and observe that if $\tau_u\geq\tau_v$ then $B|_{D_u=D_v=D}<0$. Indeed, $\partial_uf>0$, being $u$ the activator species, and $\mathrm{tr}(J_0)<0$ by the stability assumption on $p_1(\lambda)$; hence the term in brackets on the right hand side is the sum of three positive terms, from which the claim follows. On the contrary, if $\tau_u<\tau_v$, then $B|_{D_u=D_v=D}>0$ provided 
\begin{equation*}
 -\mathrm{tr}(J_0)>\frac{\tau_u+\tau_v}{\tau_u(\tau_v-\tau_u)}-\frac{\tau_u+\tau_v}{\tau_u}\partial_u f\, .
\end{equation*}
Finally, condition~\eqref{eq:inst22} no longer depends on $D$ and can thus be verified by a suitable choice of the remaining parameters.

In conclusion, we can have Turing instability also in the case of equal diffusivites, $D_u=D_v$, provided the inhibitor has a larger inertial time than the activator, $\tau_v>\tau_u$.

Let us conclude this section by considering again the case $\tau_u=\tau_v=\tau$. Because of the previous analysis we have to assume $D_u\neq D_v$, otherwise no Turing instability can develop. Then Eq.~\eqref{eq:inst21} simplifies into
\begin{equation*}
B|_{\tau_u=\tau_v=\tau}=-2 \tau \left[ (D_u+D_v) +\tau (D_v-D_u)(\partial_u f-\partial_v g)\right]\, .
\end{equation*} 
Being  $v$ the inhibitor species we have $\partial_v g<0$, hence, if $D_v>D_u$, we can conclude that the term in brackets on the right hand side is the sum of positive terms and thus $B|_{\tau_u=\tau_v=\tau}<0$. On the other hand, if $D_v<D_u$, we can have $B|_{\tau_u=\tau_v=\tau}>0$ provided that
\begin{equation*}
 \tau > \frac{D_u+D_v}{D_u-D_v}\frac{1}{\partial_u f-\partial_v g}\, .
\end{equation*}
Finally, the remaining condition~\eqref{eq:inst22} can be rewritten as
\begin{equation*}
 4\tau^2\left[(D_u+D_v) ^2+2 D_uD_v\mathrm{tr}(J_0)-4\tau^2(D_v-D_u)^2\det J_0\right]>0\, ,
\end{equation*}
and a straightforward computation allows to show that it is satisfied if
\begin{equation*}
 \tau > \frac{D_uD_v\mathrm{tr}(J_0)+\sqrt{\left[D_uD_v\mathrm{tr}(J_0)\right]^2+4(D_v^2-D_u^2)^2\det(J_0)}}{4 (D_v-D_u)^2\det(J_0)}\, .
\end{equation*}

Let us stress that in this setting, $D_v<D_u$, the conditions~\eqref{eq:inst21} and~\eqref{eq:inst22}  cannot be satisfied, hence the emergence of Turing instability is solely due to the finite propagation velocity and imposes a lower bound on the inertial time. Before introducing the model we will use to present our results, let us \textcolor{black}{emphasise two more relevant results. First, the proposed framework allows to prove the emergence of Turing instability also in an inhibitor-inhibitor system, that is $\partial_u f <0$ and $\partial_v g<0$; indeed, while Eq.~\eqref{eq:inst11} cannot hold true, Eqs.~\eqref{eq:inst21} and~\eqref{eq:inst22} can be satisfied for suitable choice of the parameters, as we will show below (see Fig.~\ref{fig:inhinh} and the associated discussion). Second, inertia-driven Turing instability cannot manifest in suitable $m$-species linear kinetic models, as described in the following remark.}

\textcolor{black}{
\begin{remark}[Kinetic linear systems]
\label{re:kinlin}
As shown in~\cite{toth} an $m$-species non-relativistic kinetic system can be expressed as a polynomial differential equation assuming mass-action for the reaction rates; however not all polynomials can be considered models of chemical reactions~\cite{Hars} because of the possible presence of negative cross terms, i.e., the abundance of a species decreases in a process where it is not involved. The absence of negative cross effects in the case of first order kinetic differential systems, prevents the Turing instability~\cite{toth}. We can prove a similar result to hold true in the relativistic framework, provided all the species have the same inertial time and the ratios of the diffusion coefficients divided by the inertial time are large enough
%; on the other hand if the latter condition is relaxed, we can exhibit a $2$-species linear kinetic system without cross inhibition terms but still experiencing a Turing instability 
(see Appendix~\ref{sec:Applinkin}).
\end{remark}}

\section{The FitzHugh-Nagumo model}
\label{sec:FHN}
\noindent
The aim of this section is to present an application of the theory hereby developed. For the sake of definitiveness, we decided to use the FitzHugh-Nagumo model~\cite{fitzhugh,nagumo,Rinzel}, but  our results go beyond the chosen model. The FitzHugh-Nagumo model (for short {\em FHN}) is a paradigmatic nonlinear system already used in the literature to study the emergence of Turing patterns~\cite{Ohta,Shoji,CarlettiNakao,siebert} as well as synchronisation phenomena~\cite{Vragovic,AQIL20121615}. \textcolor{black}{Let us observe that this model is not a kinetic one, since the $-v$ term appearing in the rate evolution for $u$, expresses a negative cross-effect~\cite{toth}, at the same time this supports our statement that Turing instability finds applications beyond the morphogenesis and chemical frameworks.} Our choice relies also on the observation that such model has been conceived in the framework of neuroscience as a schematisation of an electric impulse propagating through an axon. For this reason, we believe that it would make a suitable setting to account for a finite velocity propagation of signals and it could be interesting for future applications.

\begin{figure}[ht]
\centering
\includegraphics[scale=0.25]{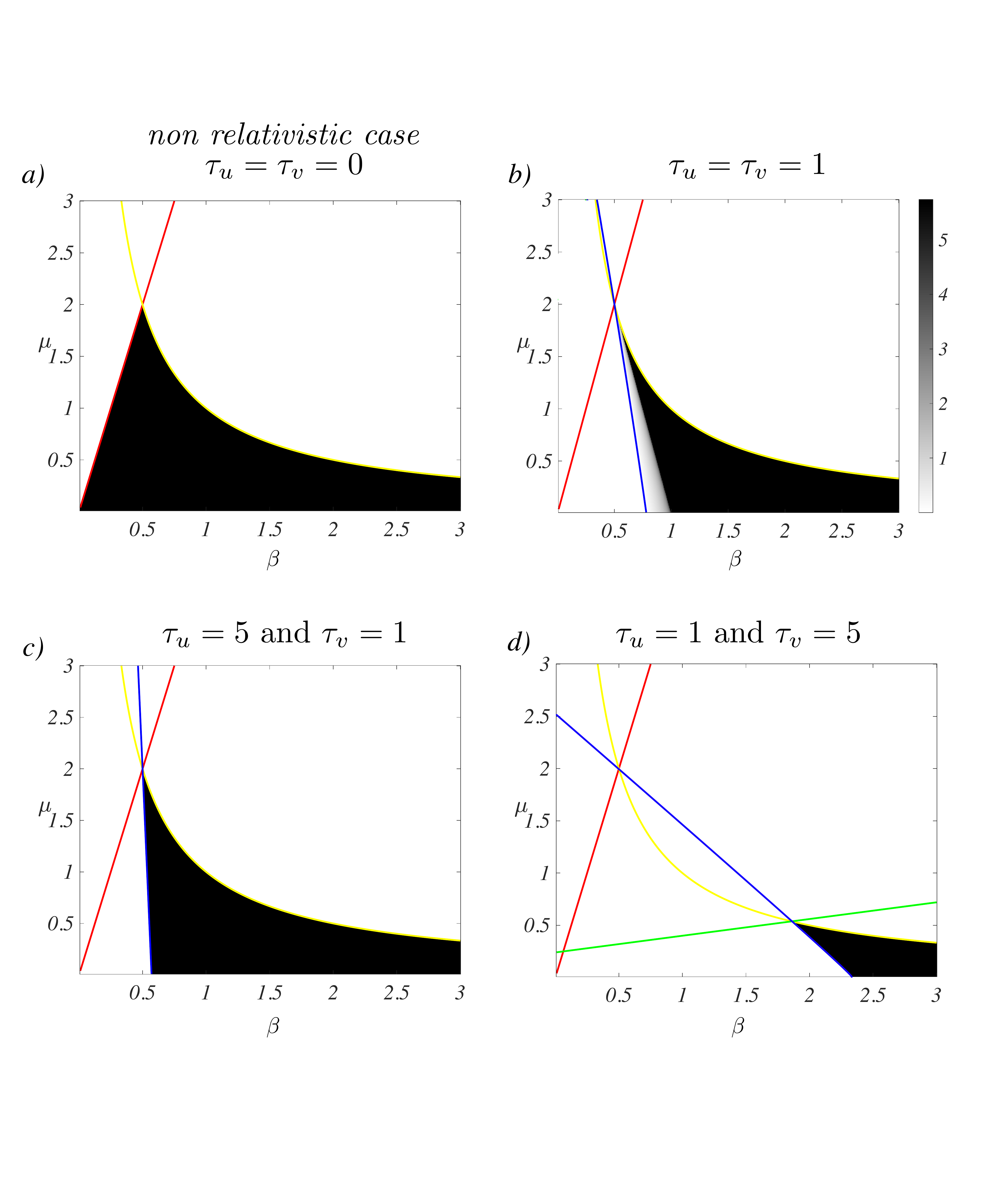}
\vspace{-1.5cm}
\caption{\textbf{\textcolor{black}{Parameter} region associated to the stability of the homogeneous solution for the {\em FHN} model.} For a fixed value of $\gamma=4$, we study the stability of the homogeneous equilibrium $(u_i,v_i)=(0,0)$, $i=1,\dots,n$, as a function of $\beta$ and $\mu$: the black regions denote stability while white ones instability. Panel a) corresponds to the classical setting, i.e., $\tau_u=\tau_v=0$, the remaining panels are associated to positive values of the inertial times, $\tau_u=\tau_v=1$ (panel b)), $\tau_u=5$ and $\tau_v=1$ (panel c)) and $\tau_u=1$ and $\tau_v=5$ (panel d)). In all the panels the red line denotes the condition $\mathrm{tr}(J_0)=0$, while $\det(J_0)=0$ is represented by the yellow one; these two lines determine the boundary of the stability region in the classical setting. Such region is shrunk in the case of positive inertial times because of the additional constraints, Eq.~\eqref{cond14} (green line) and Eq.~\eqref{cond15} (blue one). The grey shaded region in panel b), coloured according to $\ln\textcolor{black}{\tau_{\mathrm{max}}}$, is associated to a stability of the homogenous equilibrium constrained to a bound on $\tau$, see Eq.~\eqref{eq:cond14taubound}, while in the black region any positive value of $\tau$ is admissible.}
\label{fig:mualphaFHN}
\end{figure}

The {\em FHN} model can be described by the system of ODE
\begin{equation}
\label{eq:FHN}
\begin{dcases}
\frac{du}{dt}&= \mu u-u^3-v \\
\frac{dv}{dt}&= \gamma(u-\beta v) \, .
\end{dcases}
\end{equation}
where the parameters $\mu$, $\gamma$ and $\beta$ are assumed to be positive. We will hereby focus on its behaviour close to the fixed point $(u_0,v_0)=(0,0)$. The linear stability analysis ensures stability of the latter under the conditions $\mu<\gamma\beta$ and $\mu\beta<1$ (see panel a) in Fig.~\ref{fig:mualphaFHN}). Let us observe that, once such conditions are not met, the system undergoes a supercritical Hopf-Andronov bifurcation~\cite{strogatzbook}: the equilibrium point becomes unstable giving birth to a limit cycle solution. In this study we will limit ourselves to the former case, leaving the oscillating case for a future work.

Consider now $n$ identical copies of the FitzHugh-Nagumo model~\eqref{eq:FHN} interacting with each other through a diffusive-like coupling and assume to work in the Cattaneo framework presented in Section~\ref{sec:RRDnet}. The resulting model can thus be written as
\begin{equation}
\label{eq:FHNRelnet}
\begin{dcases}
\frac{du_i}{dt}+\tau_u \frac{d^2u_i}{dt^2}&=\mu u_i-u_i^3-v_i+D_u\sum_{j=1}^{n}L_{ij} u_j  \\ 
\frac{dv_i}{dt}+\tau_v \frac{d^2v_i}{dt^2}&=\gamma(u_i-\beta v_i)+D_v\sum_{j=1}^{n}L_{ij} v_j
\end{dcases}\quad \forall i=1,\dots n\, ,
\end{equation}
where $D_u$ (resp. $D_v$) is the diffusion coefficients of species $u$ (reps. $v$) and $\tau_u$ (resp. $\tau_v$) the inertial time for species $u$ (resp. $v$). The matrix $\mathbf{L}$ is the Laplace matrix describing the diffusive coupling among the FitzHugh-Nagumo systems.

\vspace{1Em}
\textbf{Remark} ({\em About the network substrate})
The possible onset of Turing instability depends both on the dynamical system as well as the network substrate via the eigenvalues  $\Lambda^{(\alpha)}$ of the associated Laplace matrix, $\mathbf{L}$. As previously stated, a discrete topology may affect the dynamics due to finite size effects. Because a cohemprensive study of such impact on Turing instability goes beyond the scope of this paper and for the sake of definitiveness, we decided to use an Erd\H{o}s-{R\'enyi} random graph~\cite{erdosreny} made of $n$ nodes and {each couple of nodes having a probability $p\in (0,1)$ to be linked}. In the following we fixed $n=30$ and $p=0.1$ and we also checked that the resulting network is connected.\\

In the rest of this Section we will present our analysis about the emergence of Turing instability in the relativistic {\em FHN} defined on networks~\eqref{eq:FHNRelnet}. Let us start by determining the \textcolor{black}{parameter} region associated to a stable homogeneous equilibrium, $(u_0,v_0)=(0,0)$. The panel a) of Fig.~\ref{fig:mualphaFHN} represents the classical case where we assume an infinite propagation velocity, namely $\tau_u=\tau_v=0$. The stability region (black) is delimited by the conditions $\mu<\gamma\beta$ (red line) and $\mu\beta<1$ (yellow line). In panel b) we report the case of equal inertial times, $\tau_u=\tau_v=1$; we can observe that the stability region (black and shades of grey) is contained in the previous one, being delimited by the same conditions and in addition by Eq.~\eqref{cond15} (blue line). As previously observed, the stability of the homogeneous solution depends on the value of $\tau_u=\tau_v=\tau$, meaning that the equilibrium loses its stability if the inertial time is too large, as shown by Eq.~\eqref{eq:cond14taubound}. The grey shaded region in panel b) has thus been coloured according to $\ln\textcolor{black}{\tau_{\mathrm{max}}}$: smaller values are associated to lighter shades of grey. On the contrary, in the black region any positive value of $\tau$ returns a stable homogeneous equilibrium (being $\textcolor{black}{\tau_{\mathrm{max}}}=\infty$). In the remaining panels of Fig.~\ref{fig:mualphaFHN} we considered different inertial times, $\tau_u=5$ and $\tau_v=1$ in panel c), and $\tau_u=1$ and $\tau_v=5$ in panel d). The stability region (black) is delimited by the same lines as before, with the exception of the case $\tau_u<\tau_v$, where an extra condition needs to be considered, i.e., Eq.~\eqref{cond14} (green line).

\begin{figure}[ht]
\centering
\includegraphics[scale=0.24]{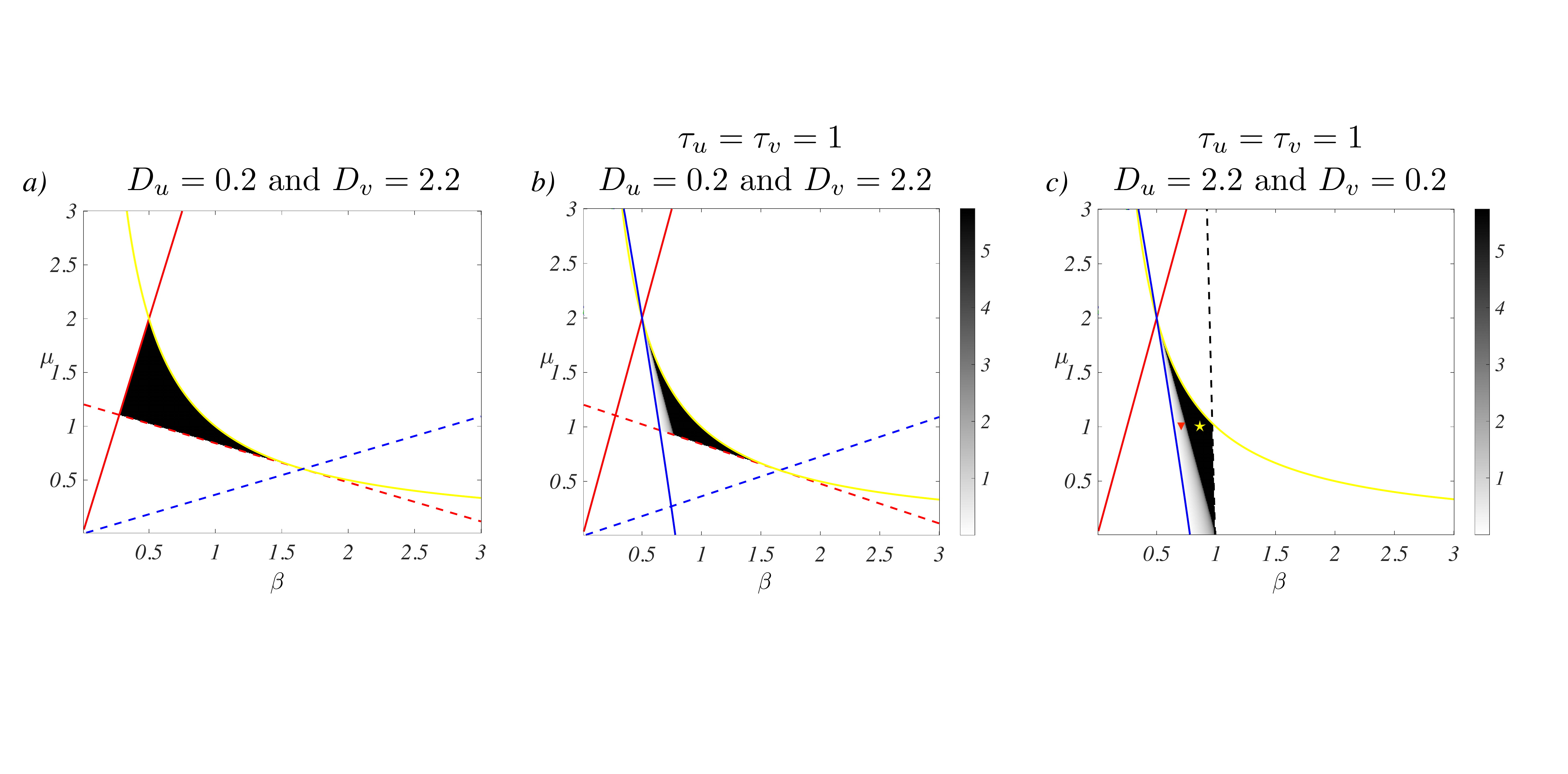}
\vspace{-1.5cm}
\caption{\textbf{\textcolor{black}{Parameter} region associated to Turing instability for the {\em FHN} model, $\tau_u=\tau_v$.} For a fixed value of $\gamma=4$, we study the onset of Turing instability (black regions) close to the homogeneous equilibrium $(u_i,v_i)=(0,0)$, $i=1,\dots,n$, as a function of $\beta$ and $\mu$. Panel a) corresponds to the classical setting, i.e., $\tau_u=\tau_v=0$, the remaining panels are associated to positive values of the inertial times, $\tau_u=\tau_v=1$. In panels a) and b) the diffusivities have been set equal to $D_u=0.2$ and $D_v=2.2$, namely the inhibitor diffuses faster than the activator. Panel c) present a completely new setting where Turing instability can develop even for a slower inhibitor, $D_u=2.2$ and $D_v=0.2$. In all the panels the red line denotes the condition $\mathrm{tr}(J_0)=0$, while $\det(J_0)=0$ is represented by the yellow one. In panels a) and b), the dashed blue line represents the condition $D_v\partial_u f+D_u\partial_v g=0$ (Eq.~\eqref{eq:inst11}), while the dashed red line the condition $(D_u\partial_v g+D_v\partial_u f)^2-4{D_uD_v} \det(J_0)=0$ (Eq.~\eqref{eq:inst12}). Together with the blue line in panel b) corresponding to Eq.~\eqref{cond15}, these lines delimitate the \textcolor{black}{parameter} region allowing for Turing instability in the case $D_u < D_v$. In panel c), corresponding to $D_u > D_v$, a similar \textcolor{black}{parameter} region is bounded by the same blue line but also by the dashed black line, namely Eq.~\eqref{eq:inst22}. The grey shaded region in panels b) and c), coloured according to $\ln\textcolor{black}{\tau_{\mathrm{max}}}$, is associated to a stability of the homogeneous equilibrium constrained to a bound on $\tau$, see Eq.~\eqref{eq:cond14taubound}, while in the black region any positive value of $\tau$ is admissible.}
\label{fig:mualphaFHNTPsametau}
\end{figure}

We are now able to study the emergence of Turing instability under the assumption $\tau_u=\tau_v=\tau$. In the panel a) of Fig.~\ref{fig:mualphaFHNTPsametau} we report the region (black) in the \textcolor{black}{parameter} space allowing for classical Turing instability to arise for a choice of the diffusivities $D_u<D_v$. Such region is contained in the one associated to a stable homogeneous solution and it is now also bounded by the conditions $D_v\partial_u f+D_u\partial_v g=0$ (dashed blue line) and  $(D_u\partial_v g+D_v\partial_u f)^2-4{D_uD_v} \det(J_0)=0$ (dashed red line). The same values of the parameters are used in panel b) assuming now the inertial times to be positive, $\tau_u=\tau_v=1$; the Turing region (black and shades of grey) is smaller, as it is also delimited by the condition Eq.~\eqref{cond15} (blue line). Once again, the shades of grey represent the values of $~\ln\textcolor{black}{\tau_{\mathrm{max}}}$ to ensure the stability of the homogeneous solution (see Eq.~\eqref{eq:cond14taubound}). Finally in panel c) we report the analysis of a setting for which classical Turing instability can never emerge because the inhibitor diffuses slower than the activator,  $D_u =2.2 > D_v=0.2$. {The instability being determined by the inertial time $\tau>0$}, we named it {\em inertia-driven instability}. The Turing region (black and shades of grey) is now delimited also by the condition Eq.~\eqref{eq:inst22}, where again the shades of grey represent the values of $~\ln\textcolor{black}{\tau_{\mathrm{max}}}$.

The impact of $\textcolor{black}{\tau_{\mathrm{max}}}$ can be appreciated in Fig.~\ref{fig:FHNTPsametau} where we report {for few generic sets of parameters} the {\em dispersion relation}, $\lambda_\alpha$, as a function of $\Lambda^{(\alpha)}$, the eigenvalues of the Laplace matrix, $\mathbf{L}$. In panel a) we show the dispersion relation for the choice $\tau_u=\tau_v=1$ and $(\beta,\mu)=(0.8,1.0)$, lying in the Turing instability region  (yellow star in the panel c) of Fig.~\ref{fig:mualphaFHNTPsametau}). We can observe that the homogeneous equilibrium is stable (the dispersion relation is negative for $\Lambda^{(1)}=0$) but it turns out to be unstable under heterogeneous perturbations (there exist $\alpha>1$ (red dots) for which $\lambda_\alpha>0$) and synchronised oscillatory patterns emerge ({see inset where we report $u_i(t)$ vs $t$~\footnote{Throughout this work the numerical simulations have been performed using a $4$-th order Runge-Kutta scheme implemented in Matlab~\cite{MATLAB2021}; the code is available upon request to the corresponding author. The initial conditions have been realised by drawing uniformly random perturbations $\delta$-close to the homogeneous equilibrium and the simulation time has been taken of the order of $-\log \delta / \max_\alpha \Re\lambda_\alpha$. Indeed according to the ansatz $\hat{u}_\alpha \sim e^{\lambda_\alpha t}$ and $\hat{u}_\alpha \sim e^{\lambda_\alpha t}$, this is the time necessary to (possibly) increase the $\delta$-perturbation up to a macroscopic size. In the rest of the work we set $\delta=10^{-2}$, small enough to discriminate between the onset of the instability using a reasonable simulation time for the values of $ \max_\alpha \Re\lambda_\alpha$ we are dealing with.}), {indeed, we are in presence of an oscillatory Turing instability because $\rho_\alpha>0$ (data not shown)}. Panel b) ($\tau_u=\tau_v=2.2$ and $(\beta,\mu)=(0.7,1.0)$, red triangle in the panel c) of Fig.~\ref{fig:mualphaFHNTPsametau}) corresponds to a similar behaviour, being the parameters still in the Turing region but conditioned to the value of $\textcolor{black}{\tau_{\mathrm{max}}}$; the dispersion relation assumes positive values but the homogeneous equilibrium is weakly stable, the dispersion relation is negative but very close to $0$ for $\Lambda^{(1)}=0$, being $\tau_u=\tau_v=2.2$ close to $\textcolor{black}{\tau_{\mathrm{max}}}\sim 2.31$; {again, an oscillating synchronous behaviour is observed (inset)}. In panel c) we used the same parameters $(\beta,\mu)$ but we increased the inertial times beyond the critical values, $\tau_u=\tau_v=3.5>\textcolor{black}{\tau_{\mathrm{max}}}$, and indeed the homogeneous equilibrium is unstable, the dispersion relation is positive for $\Lambda^{(1)}=0$. Again synchronised oscillatory patterns emerge ({inset}), they are indistinguishable from the ones one can obtain from the setting presented in panels a) and b) but they are not the result of Turing instability.

\begin{figure}[ht]
\centering
\includegraphics[scale=0.25]{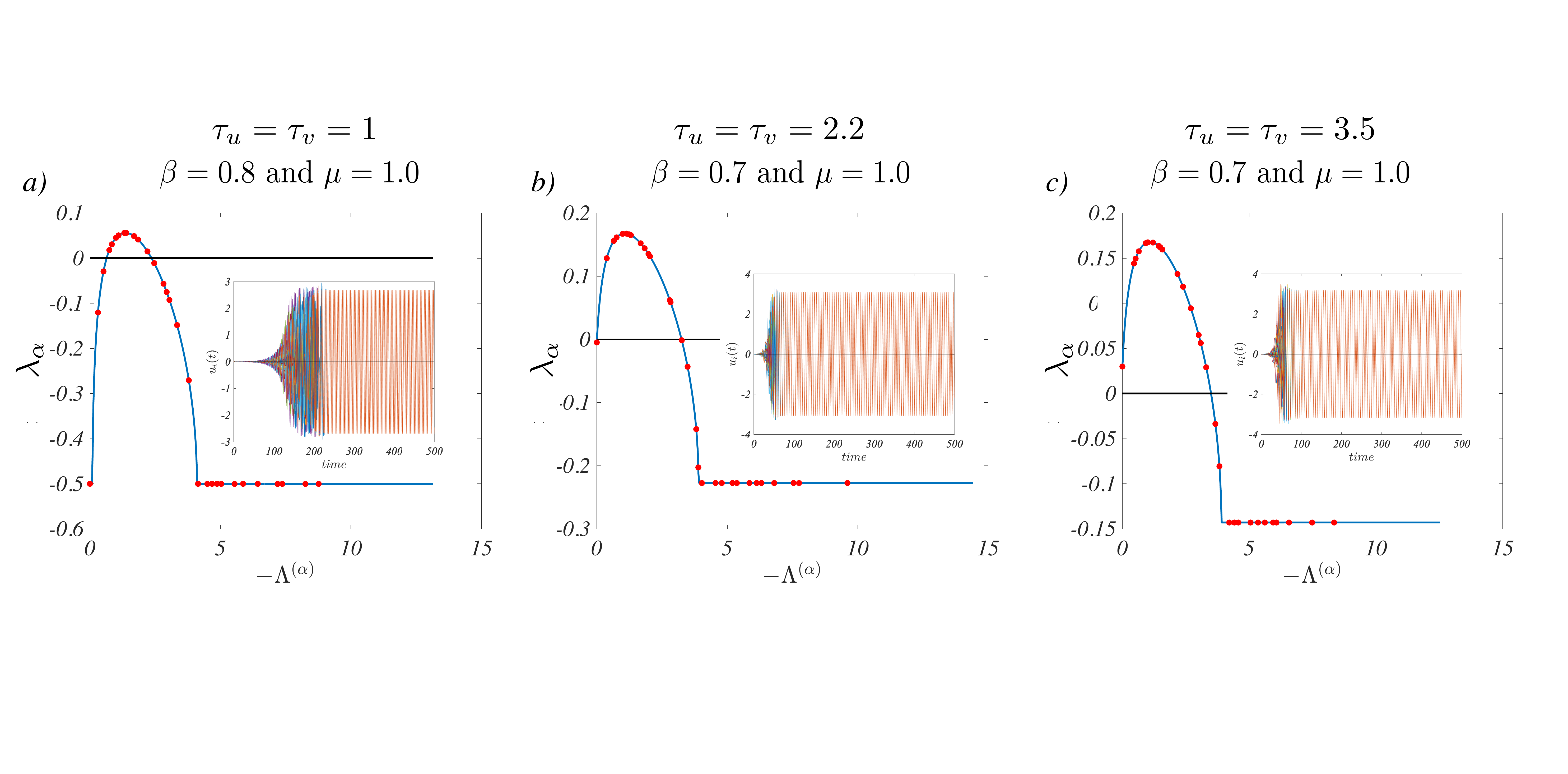}
\vspace{-1.5cm}
\caption{\textbf{Dispersion relation {and pattern} for the {\em FHN} model, $\tau_u=\tau_v$.} For fixed values of $\gamma=4$, $D_u=2.2$, $D_v=0.2$ and two couples $(\beta,\mu)$ we show {in the main panels} the dispersion relation, $\lambda_\alpha$ as a function of $\Lambda^{(\alpha)}$. Panel a) corresponds to the choice $\tau_u=\tau_v=1$ and $(\beta,\mu)=(0.8,1.0)$ (yellow star in the panel c) of Fig.~\ref{fig:mualphaFHNTPsametau}), lying the Turing instability region and indeed the dispersion relation assumes positive values (red dots lying on the positive part of the blue curve). The homogeneous equilibrium is stable (the dispersion relation is negative for $\Lambda^{(1)}=0$), but it turns out to be unstable under heterogeneous perturbations and synchronised oscillatory patterns emerge {(inset), indeed the critical root has positive imaginary part, $\rho_\alpha>0$} (conditions~\eqref{eq:inst21} and~\eqref{eq:inst22} are satisfied). In panel b) we fix $\tau_u=\tau_v=2.2$ and $(\beta,\mu)=(0.7,1.0)$ (red triangle in the panel c) of Fig.~\ref{fig:mualphaFHNTPsametau}), still in the Turing region but conditioned to the value of $\textcolor{black}{\tau_{\mathrm{max}}}$. The behaviour is similar to the one reported in panel a) but now the homogeneous equilibrium is weakly stable, the dispersion relation is negative but very close to $0$ for $\Lambda^{(1)}=0$, indeed for these values of the parameters we have $\textcolor{black}{\tau_{\mathrm{max}}}\sim 2.31$. {Again, an oscillatory behaviour is obtained (inset) associated to $\rho_\alpha>0$}. In panel c) we used the same parameters $(\beta,\mu)$ but we increased $\tau_u=\tau_v=3.5>\textcolor{black}{\tau_{\mathrm{max}}}$ and indeed the homogeneous equilibrium is unstable, the dispersion relation is positive for $\Lambda^{(1)}=0$. Again synchronised oscillatory patterns emerge {(inset)}, they are indistinguishable from the ones one could obtain with the parameters used in panels a) and b) but they are not the result of Turing instability.}
\label{fig:FHNTPsametau}
\end{figure}

We can now consider the more general case of different inertial times and show the onset of Turing instability for a choice of the diffusivities that cannot allow for the classical Turing phenomenon, notably because the activator can diffuse faster than the inhibitor. For this reason we hereby stress again that {\em inertia-driven instability} should be a suitable name for such phenomena.

In Fig.~\ref{fig:mualphaFHNTPdifftau1} we report the region (black) in the \textcolor{black}{parameter} space $(\beta,\mu)$ allowing for Turing instability under the assumptions $\tau_u\neq \tau_v$ and $D_u\geq D_v$. Such region is contained in the region associated to a stable homogeneous solution (see panels c) and d) of Fig.~\ref{fig:mualphaFHN}) and delimited in addition by the conditions~\eqref{cond14} (green line),~\eqref{cond15} (magenta line) and~\eqref{eq:inst22} (dashed black line). We can observe that for all the choices of the inertial times and diffusion constants ($\tau_u=5$, $\tau_v=1$, $D_u=2.2$ and $D_v=0.2$ panel a), $\tau_u=1$, $\tau_v=5$, $D_u=2.2$ and $D_v=0.2$ panel b) and $\tau_u=1$, $\tau_v=5$, $D_u=D_v=2.2$ panel c)) there are always parameters $(\beta,\mu)$ allowing Turing instability to occur. {One can show that in all the presented cases we are dealing with a Turing oscillatory instability, driven by the inertial times in the panels b) and c). We report in Fig.~\ref{fig:FHNTPdifftau1} two generic dispersion relations for the latter settings.} In both cases we can appreciate the fact that the aspatial solution is stable, indeed $\lambda_1<0$, while there are $\alpha>1$ (red dots) for which $\lambda_\alpha>0$, testifying the instability of such equilibrium once subjected to heterogeneous perturbations and resulting into synchronised oscillatory patterns ({see panels c) and f)) associated with a positive $\rho_\alpha$ (see panels b) and e))}.

\begin{figure}[ht]
\centering
\includegraphics[scale=0.25]{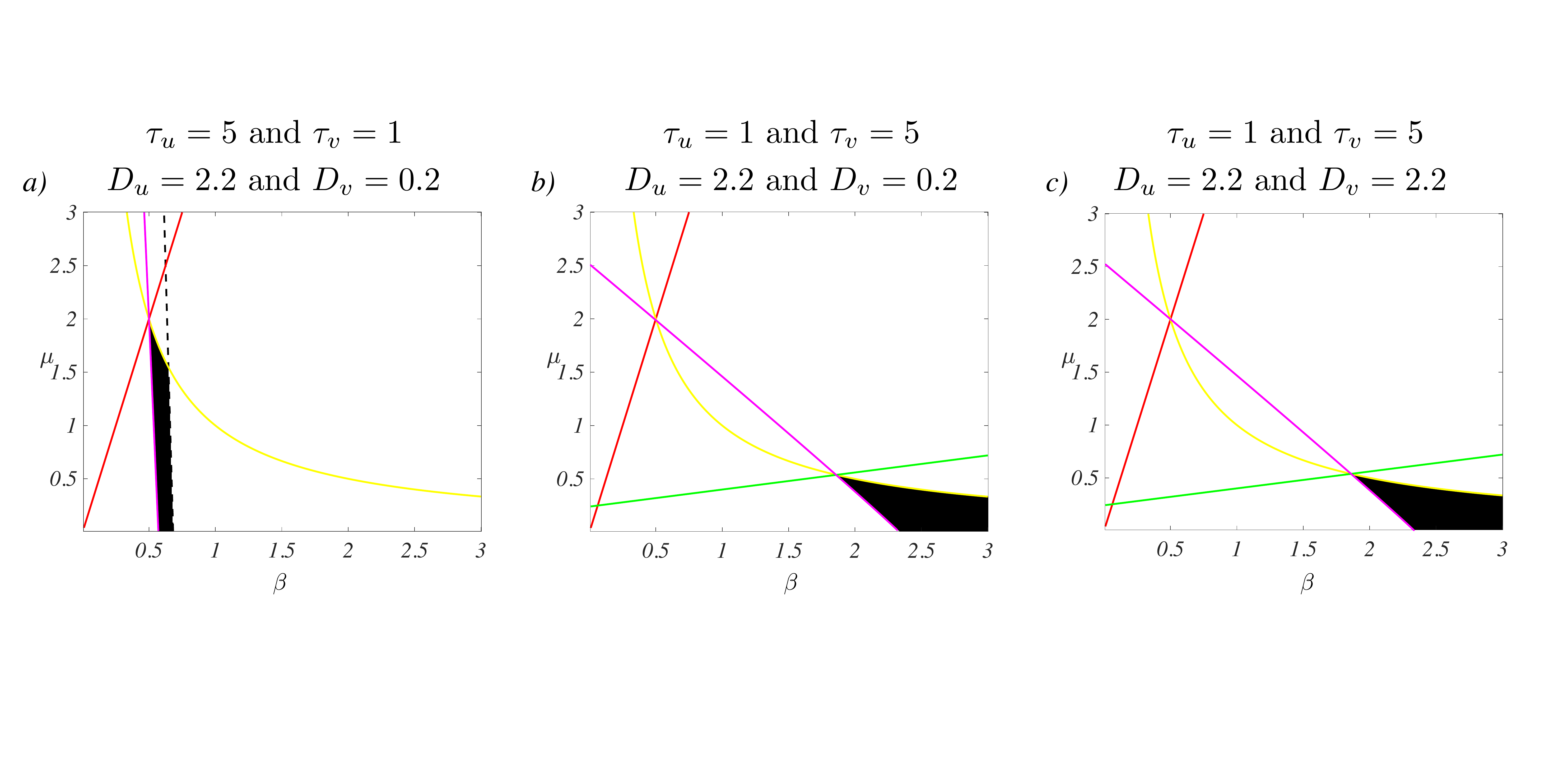}
\vspace{-1.5cm}
\caption{\textbf{\textcolor{black}{Parameter} region associated to the inertia-driven instability for the {\em FHN} model.} For a fixed value of $\gamma=4$, we study the onset of Turing instability (black regions) close to the homogeneous equilibrium $(u_i,v_i)=(0,0)$, $i=1,\dots,n$, as a function of $\beta$ and $\mu$ and driven by the inertial times, $\tau_u\neq \tau_v$. Indeed we assume $D_u \geq D_v$, resulting in a setting where classical Turing instability cannot emerge. Panel a) corresponds to the setting, $\tau_u=5$ and $\tau_v=1$, $D_u=2.2$ and $D_v=0.2$. In panel b) we use the same diffusivities while the inertial times are exchanged, i.e., $\tau_u=1$ and $\tau_v=5$. Panel c) reports result for $D_u=D_v=2.2$ and $\tau_u=1$ and $\tau_v=5$. In all the panels the red line denotes the condition $\mathrm{tr}(J_0)=0$, while $\det(J_0)=0$ is represented by the yellow one. The green line represents condition~\eqref{cond14}, while the magenta one represents condition~\eqref{cond15}; once present, the dashed black line stands for Eq.~\eqref{eq:inst22}.}
\label{fig:mualphaFHNTPdifftau1}
\end{figure}

\begin{figure}[ht]
\centering
\includegraphics[scale=0.25]{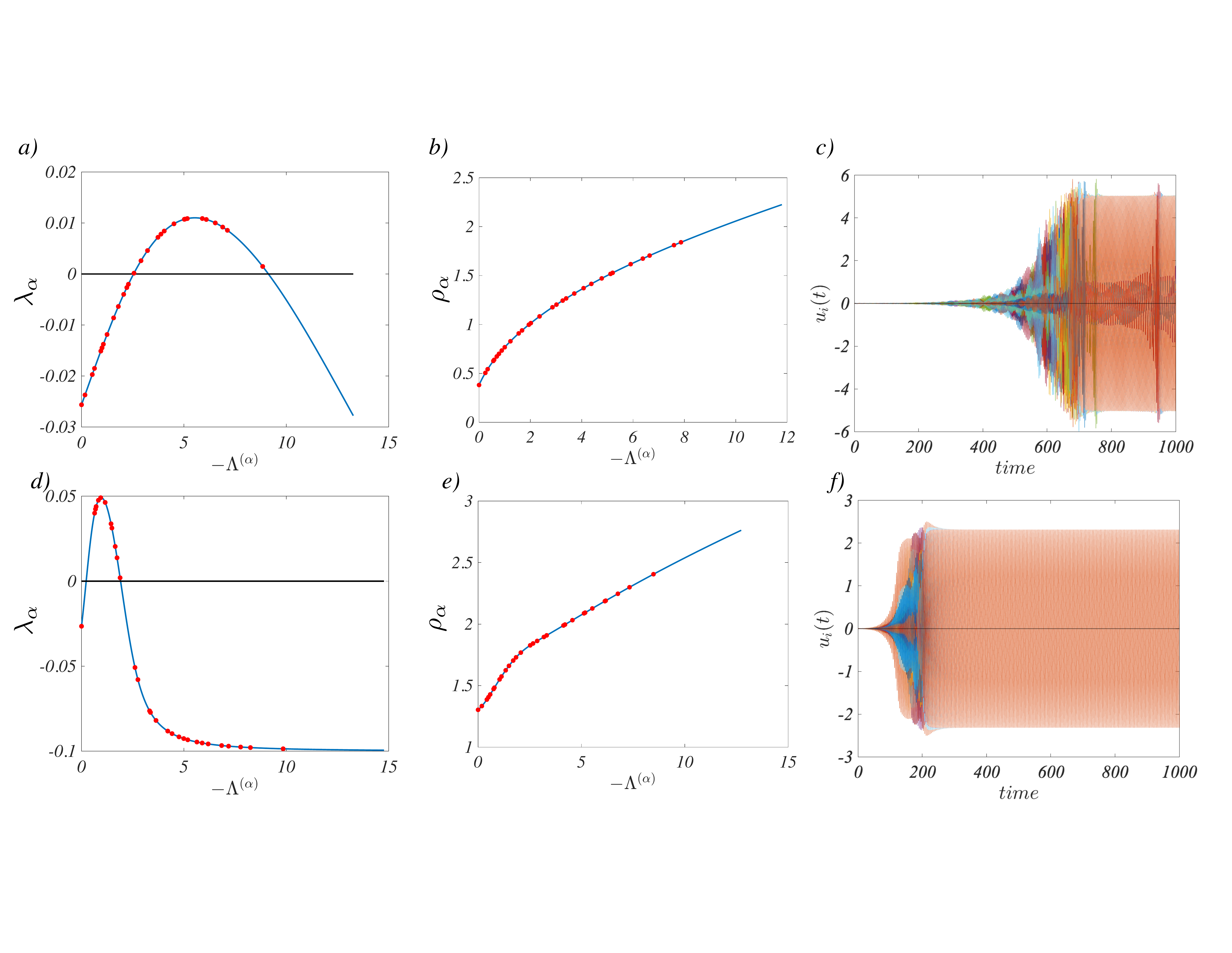}
\vspace{-1.5cm}
\caption{\textbf{Dispersion relation {and pattern} for the {\em FHN} model in the case $\tau_u\neq \tau_v$.} {For a fixed value of $\gamma=4$ and two couples $(\beta,\mu)$ and $(\tau_u,\tau_v)$ we show the dispersion relation (panels a) and d), $\lambda_\alpha$ as a function $\Lambda^{(\alpha)}$, the imaginary part of the root with the largest real part, $\rho_\alpha$ (panels b) and e)), and the time evolution of the solutions $u_i(t)$ (panels c) and f)). The top three panels correspond to the choice $\tau_u=5$ and $\tau_v=1$, $(\beta,\mu)=(0.6,1.0)$, and $D_u=2.2$ and $D_v=0.2$, lying the inertia-driven instability region (see panel a) Fig.~\ref{fig:mualphaFHNTPdifftau1}), indeed the conditions~\eqref{eq:inst21} and~\eqref{eq:inst22} hold true. While the bottom three panels are associated to $\tau_u=1$ and $\tau_v=5$, $(\beta,\mu)=(2.5,0.18)$ and with equal diffusivity $D_u=D_v=2.2$; these values are still in the inertia-driven region (see panel c) Fig.~\ref{fig:mualphaFHNTPdifftau1}) and the conditions~\eqref{eq:inst21} and~\eqref{eq:inst22} are satisfied. In both cases the aspatial equilibrium is stable ($\lambda_1<0$), but it turns out to be unstable under heterogeneous perturbations and synchronised oscillatory patterns emerge. Being $\rho_\alpha>0$ we are in presence of a Turing-wave instability driven by the inertial times.}}
\label{fig:FHNTPdifftau1}
\end{figure}

The bifurcation regions reported in Fig.~\ref{fig:mualphaFHNTPdifftau2} correspond to a parameters setting for which Turing instability could emerge because the inhibitor diffuses faster than the activator, {$D_v>D_u$, even without the presence of positive inertial times}, indeed the conditions~\eqref{eq:inst11} and~\eqref{eq:inst12} are satisfied. However, the resulting patterns {and dispersion relations (see Fig.~\ref{fig:FHNTPdifftau2})} are quite different in the relativistic case, $\tau_u>0$ and $\tau_v>0$. {The top three panels refer to a generic point in the Turing region (see panel b) Fig.~\ref{fig:mualphaFHNTPdifftau2}), here $\gamma=4$, $\beta=0.9$, $\mu=1.0$, $\tau_u=1$, $\tau_v=2$, $D_u=0.2$ and $D_v=2.2$; we can clearly appreciate that Turing instability is at play. Indeed, the aspatial equilibrium is stable, $\lambda_1<0$, and there are modes $\alpha>1$ for which the dispersion relation is positive, $\lambda_\alpha>0$ (panel a)); moreover, such unstable modes are real, being their imaginary part  zero, $\rho_\alpha=0$ (panel b)). One should thus expect the system to settle into stationary patterns, but that is not the case (panel c)); in fact, the solution departs from the homogeneous equilibrium and it spends a transient time (much longer that the initial period needed to depart from the equilibrium) oscillating with very small amplitudes around different values, only after this phase the amplitudes increase and a wave develops. Observe also that each node oscillates about a different average value, which is not the case in the oscillating patterns shown before. The bottom three panels correspond to a generic point still in the Turing region (see panel a) Fig.~\ref{fig:mualphaFHNTPdifftau2}), with $\gamma=4$, $\beta = 0.7$, $\mu=1.15$, $\tau_u=5$, $\tau_v=1$, $D_u=0.2$ and $D_v=2.2$ (conditions~\eqref{eq:inst11} and~\eqref{eq:inst12} hold true). The dispersion relation and its imaginary part behave similarly to the previous case, however now the solution diverging from the unstable homogeneous equilibrium settles onto a stationary heterogeneous equilibrium (panel f)), as one should expect from the classical, i.e., non-relativistic, Turing instability.}

To the best of our knowledge, this is a {remarkable phenomenon} that should be taken into account in the problem of patterns prediction~\cite{subram}. Indeed, since the seminal paper by A. Turing~\cite{Turing}, scholars are aware of the existence of stationary Turing patterns, often associated to a real dispersion relation, and of oscillatory Turing patters resulting from a Turing wave instability. The use of discrete substrates such as networks questioned this dichotomy and a rule of thumb seems to apply~\cite{universal_route}: oscillatory patterns develop if the most unstable mode has a large imaginary part, $\rho_\alpha \gg \lambda_\alpha$. The last example goes in the opposite direction because here $\lambda_\alpha > \rho_\alpha=0$ {and the system can exhibit stationary patterns as well as waves,} recalling that the final patterns are initiated by the linear behaviour, but rather shaped by the nonlinear character of the system.

\begin{figure}[ht]
\centering
\includegraphics[scale=0.25]{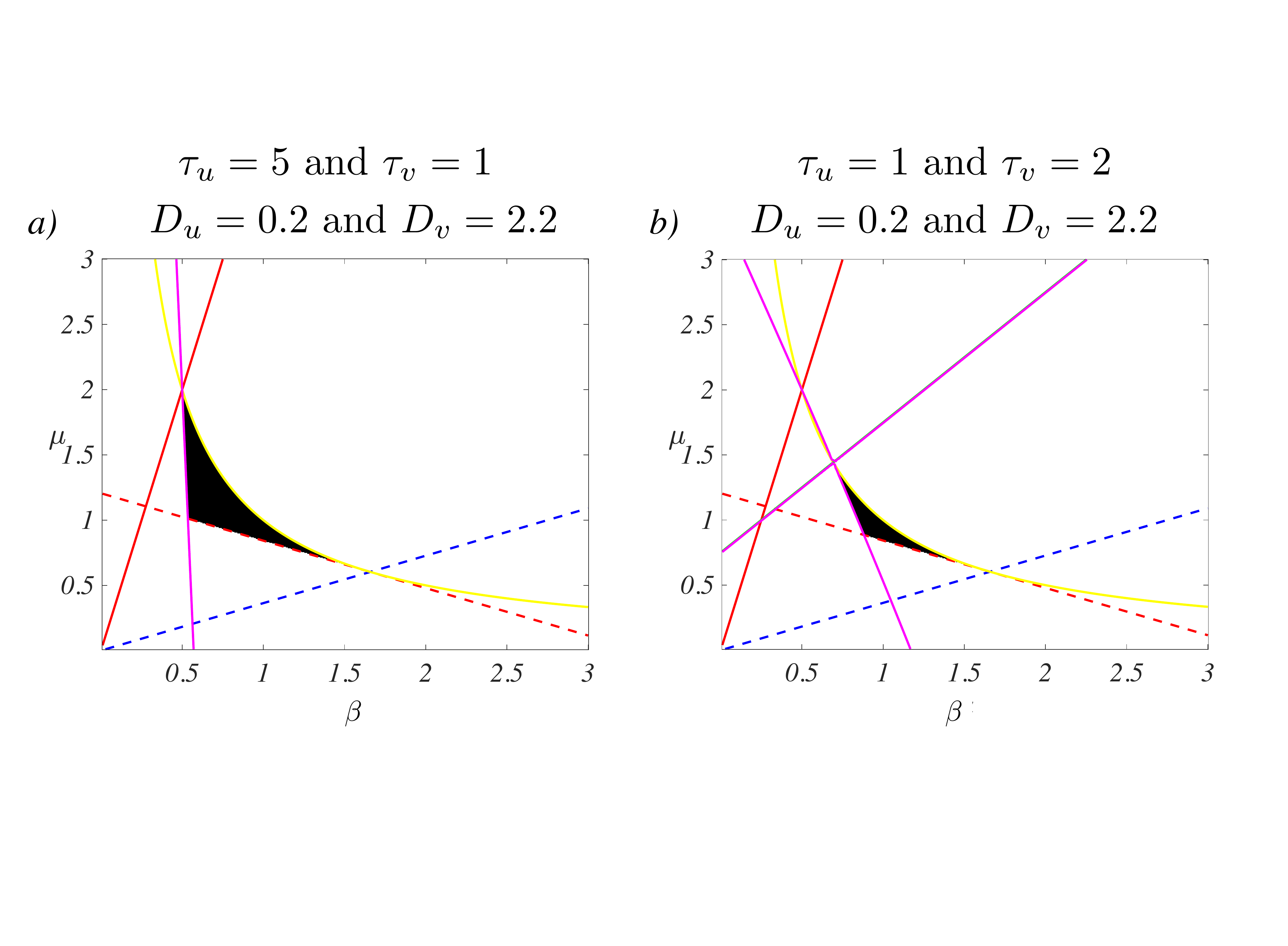}
\vspace{-1.5cm}
\caption{\textbf{\textcolor{black}{Parameter} region associated to Turing instability for the {\em FHN} model, $D_u<D_v$.} For a fixed value of $\gamma=4$, we study the onset of Turing instability (black regions) close to the homogeneous equilibrium $(u_i,v_i)=(0,0)$, $i=1,\dots,n$, as a function of $\beta$ and $\mu$ and different choices of inertial times, $\tau_u$ and $\tau_v$, and of the diffusivities, $D_u$ and $D_v$, in a setting where classical Turing instability could emerge because $D_u< D_v$. Panel a) corresponds to the setting, $\tau_u=5$ and $\tau_v=1$, $D_u=0.2$ and $D_v=2.2$, while panel b) shows results with the same diffusivities but $\tau_u=1$ and $\tau_v=2$.  In all the panels the red line denotes the condition $\mathrm{tr}(J_0)=0$, while $\det(J_0)=0$ is represented by the yellow one. The magenta line denotes the condition~\eqref{cond15}. The dashed blue line represents the condition $D_v\partial_u f+D_u\partial_v g=0$ (Eq.~\eqref{eq:inst11}), while the dashed red line the condition $(D_u\partial_v g+D_v\partial_u f)^2-4{D_uD_v} \det(J_0)=0$ (Eq.~\eqref{eq:inst12}).}
\label{fig:mualphaFHNTPdifftau2}
\end{figure}

\begin{figure}[ht]
\centering
\includegraphics[scale=0.25]{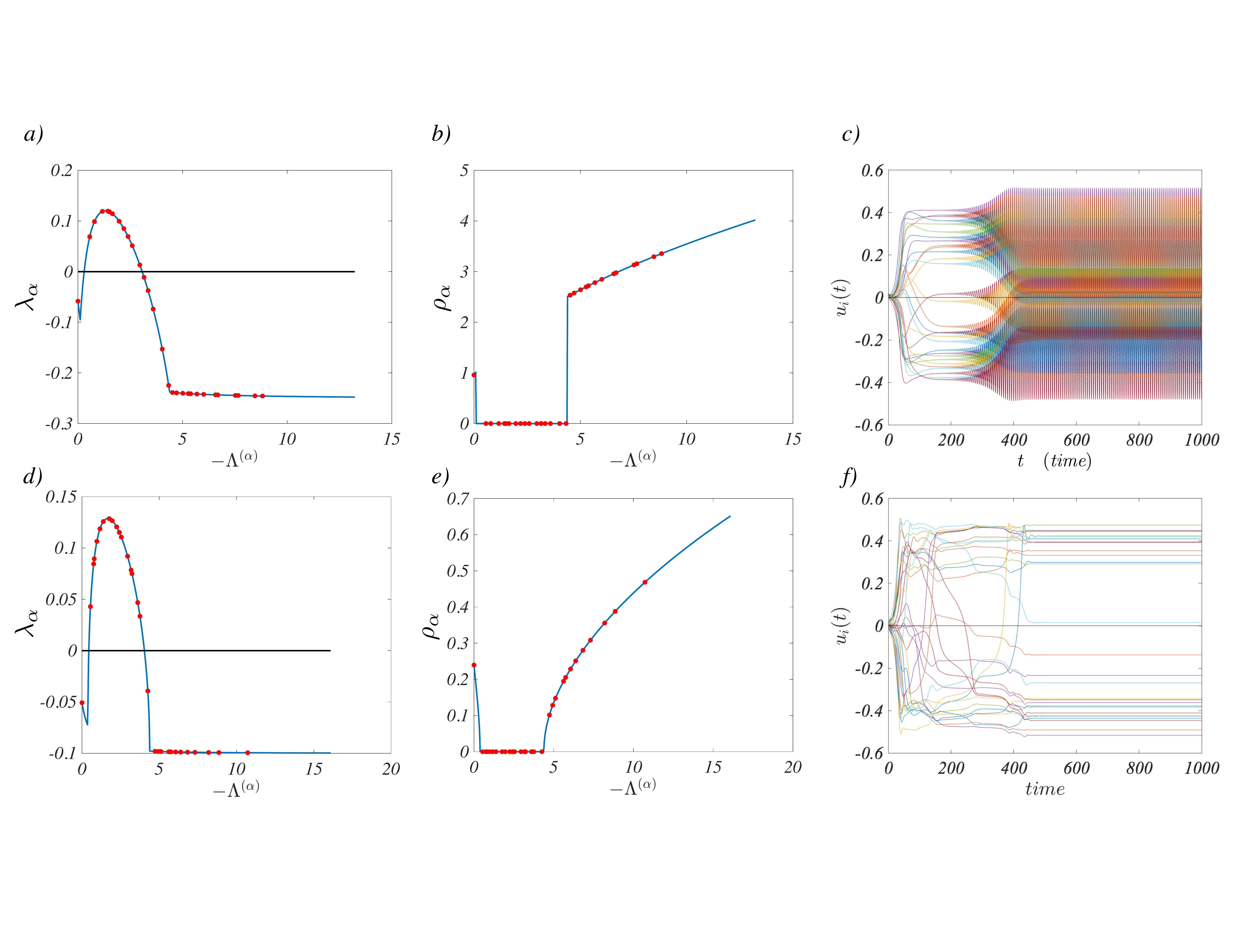}
\vspace{-1.5cm}
\caption{\textbf{Dispersion relation and pattern for the {\em FHN} model, $\tau_u\neq\tau_v$.}  {For a fixed value of $\gamma=4$ and two couples $(\beta,\mu)$ and $(\tau_u,\tau_v)$ we show the dispersion relation (panels a) and d)), $\lambda_\alpha$ as a function $\Lambda^{(\alpha)}$, the imaginary part of the root with the largest real part, $\rho_\alpha$ (panels b) and e)), and the time evolution of the solutions $u_i(t)$ (panels c) and f)). The top three panels correspond to the choice  $\tau_u=1$, $\tau_v=2$, $\beta=0.9$, $\mu=1.0$, $D_u=0.2$ and $D_v=2.2$, associated to Turing instability (see panel b) Fig.~\ref{fig:mualphaFHNTPdifftau2}); the conditions~\eqref{eq:inst11} and~\eqref{eq:inst12} hold true. While in the bottom three panels we invert the sizes of the inertial times, $\tau_u=5$ and $\tau_v=1$,  and the remaining parameters are $(\beta,\mu)=(0.7,1.15)$, $D_u=0.2$ and $D_v=2.2$, still in the Turing region (see panel a) Fig.~\ref{fig:mualphaFHNTPdifftau2}) and the conditions~\eqref{eq:inst11} and~\eqref{eq:inst12} are satisfied. In both cases the aspatial equilibrium is stable ($\lambda_1<0$), but it turns out to be unstable under heterogeneous perturbations and synchronised oscillatory patterns can emerge. Let us observe that $\rho_\alpha$ vanished on an interval containing all the unstable modes, $-\Lambda^{(\alpha)}$, and it is positive elsewhere, i.e. in correspondence to decaying modes, the obtained instability possesses thus both the characteristic of a Turing instability and a Turing-wave. Indeed the pattern associated to the first set of parameters (top panels) keeps oscillating after a transient time (panel c)) while the pattern resulting from the second set of parameters (bottom panels) settle onto a stationary solution (panel f)).}}
\label{fig:FHNTPdifftau2}
\end{figure}

\section{Discussion}
\label{sec:discuss}
\noindent
In this work we have improved the Cattaneo framework of relativistic reaction-diffusion systems to allow for complex network substrates. We have thus analytically studied the conditions for the emergence of Turing instability{, stationary or wave-like,} for hyperbolic reaction-diffusion networked systems. The introduction of the inertial times removes the unphysical assumption of infinite propagation velocity {and, more importantly this new framework allows for Turing patterns to emerge also for \textcolor{black}{parameter} values for which classical, i.e., non-relativistic, Turing instability cannot arise, e.g., once the activator diffuses faster than the inhibitor or even in the case of {\em inhibitor-inhibitor} systems.}

{To support the last claim, let us consider a generic quadratic \textcolor{black}{Lotka\,-Volterra system~\cite{lotka1920,volterra1926} involving two species}, namely $f(u,v)=u(a_1-b_1 u+c_1 v)$ and $g(u,v)=v(a_2-b_2 v-c_2 u)$, where all the parameters are positive numbers, and let us consider its relativistic networked extension:
\begin{equation}
\begin{dcases}
\frac{du_i}{dt}+\tau_u \frac{d^2u_i}{dt^2}&= u_i(a_1-b_1 u_i+c_1 v_i)+D_u\sum_{j=1}^{n}L_{ij} u_j  \\
\frac{dv_i}{dt}+\tau_v \frac{d^2v_i}{dt^2}&= v_i(a_2-b_2 v_i-c_2 u_i)+D_v\sum_{j=1}^{n}L_{ij} v_j 
\end{dcases}\quad\forall i=1,\dots, n\, .
\label{eq:inhiinhi}
\end{equation}
The homogenous nontrivial equilibrium is $u_0 = (c_1a_2+a_1b_2)/(c_2c_1+b_2b_1)$ and $v_0 = (a_2b_1-c_2a_1)/(c_2c_1+b_2b_1)$, and one can easily show that $\partial_u f = -b_1 u_0<0$ and $\partial_v g = -b_2 v_0<0$, provided $u_0>0$ and $v_0>0$ as we will hereby assume; we are hence dealing with an inhibitor-inhibitor system. Contrary to the classical setting where Turing pattern are not allowed for, in the relativistic framework parameters can be chosen in such a way that the above system exhibits an inertia-driven instability resulting in an oscillatory behaviour (see Fig.~\ref{fig:inhinh}).}
\begin{figure}[ht]
\centering
\includegraphics[scale=0.25]{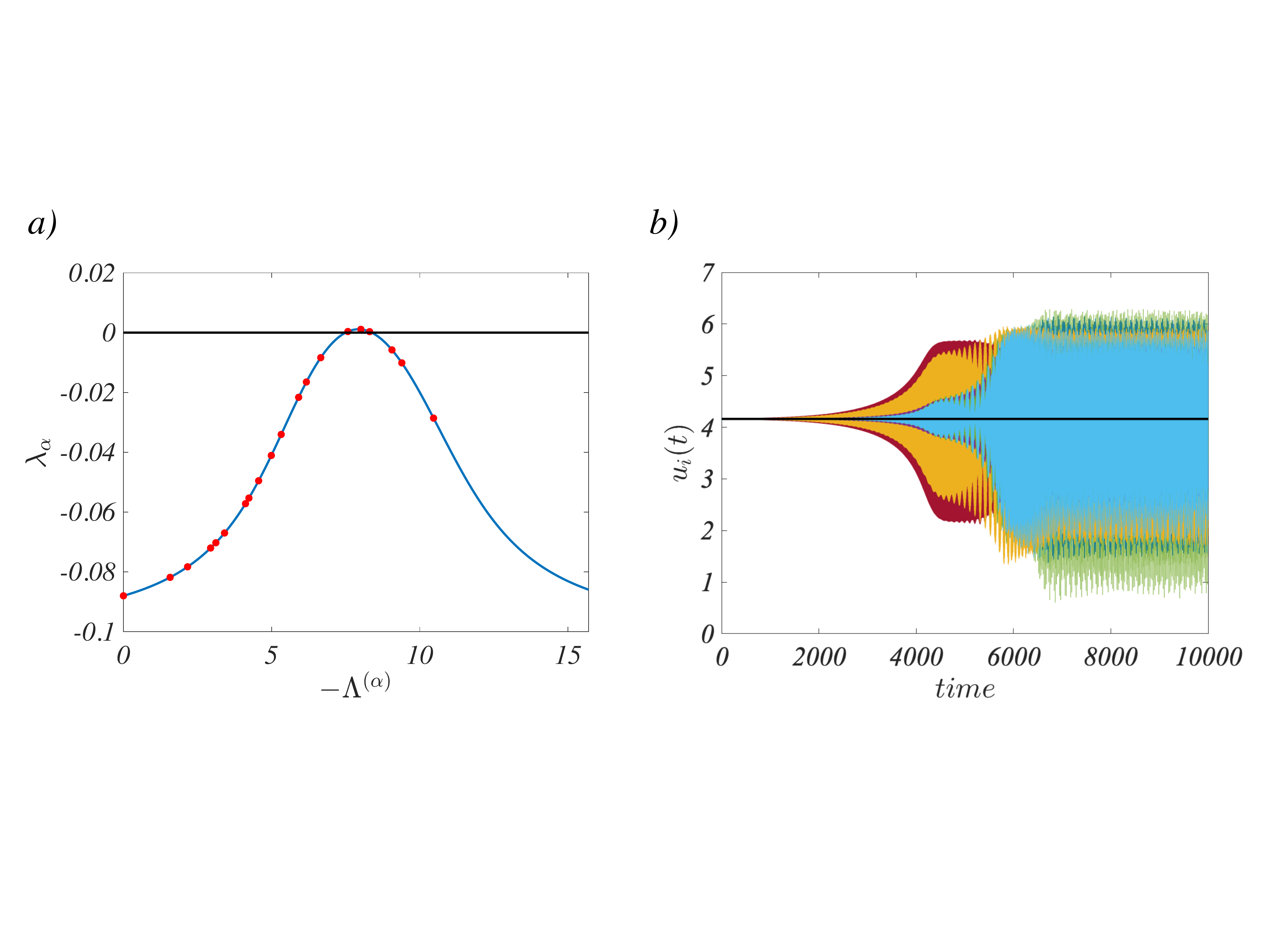}
\vspace{-1.5cm}
\caption{\textbf{Dispersion relation and pattern for the {\em inhibitor - inhibitor} model.}  {We show the dispersion relation (panel a)), $\lambda_\alpha$ as a function $\Lambda^{(\alpha)}$, and the time evolution of the solutions $u_i(t)$ (panel b)) for the system~\eqref{eq:inhiinhi}, for the choice of parameters: $a_1 = 4$, $a_2 = 3$, $b_1 = 2$, $b_2 = 1$, $c_1 = -4.7$, $c_2 = 0.5$, $D_u = 0.1$, $D_v = 5.2$, $\tau_u = 1$ and $\tau_v= 5$. We can observe that the homogeneous equilibrium is stable ($\lambda_1<0$), but it turns out to be unstable under heterogeneous perturbations (red dots in the panel a)) and oscillatory patterns can emerge (panel b)). The imaginary part of the largest roots is positive (data not shown) and we are thus dealing with an inertia-driven wave-instability.}}
\label{fig:inhinh}
\end{figure}

We have shown that the stability of the homogeneous solution is conditional to the inertial time common to both species. There exists a threshold, $\textcolor{black}{\tau_{\mathrm{max}}}$, that, if exceeded, returns an unstable homogeneous solution: the system exhibits patterns but they are not ascribed to a Turing instability; let us observe that the latter are indistinguishable from the ones emerging following the Turing mechanism.
 Interestingly enough, such threshold depends on the model parameters and it can become arbitrarily large for a specific range of the latter; in such case, the homogeneous equilibrium is always stable (with respect to the inertial time). For generic values of the inertial times, $\tau_u\neq \tau_v$, we have proved that Turing instability can set up both for the inhibitor diffusing faster than the activator, $D_v>D_u$, as it occurs in the classical setting, but also in the complementary regime, i.e., $D_v\leq D_u$, which is forbidden in the absence of inertial time. {Even more striking, Turing patterns can emerge also in the case of inhibitor-inhibitor systems. The framework we propose allows to relax the severe parameters conditions for the patterns onset and thus provide new insight into the fine tuning problem~\cite{HG2021,KGMK2021}. Hence, existing experiments could also be read with this novel perspective and analysed in the proposed framework.}

We have complement our general analytical results with a numerical study of the FitzHugh-Nagumo model extended to the framework of hyperbolic reaction-diffusion networked systems. We have found stationary patterns as well as synchronised oscillatory ones; we have also found an interesting class of solutions where the system spends a transient time into a stationary-like regime but then it evolves into an oscillatory one. This example raises relevant questions about the prediction of the patterns following a Turing instability, which is, up to now, an open problem \cite{subram}.

The investigation discussed in this paper could be further extended in several directions. Previous studies have shown that different kinds of networks, such as directed~\cite{Asllani2014NC} or non-normal ones~\cite{malbor_teo, jtb}, extend the conditions for the emergence of patterns and allow for a richer spectrum of instabilities. Moreover, it has been shown that an instability similar to the Turing mechanism can be obtained by perturbing a stable limit cycle~\cite{chall} and that non-normal networks further enhance such instability~\cite{entropy}. Given the oscillatory behaviours of neurones~\cite{fitzhugh, nagumo} and the non-normal nature of neural networks~\cite{malbor_teo}, an extension towards such direction would open the way to interesting new results and applications.

\vspace{1cm}

%\ethics{Insert ethics text here.}

\noindent\textbf{Acknowledgements}\\
\noindent The authors are grateful to Duccio Fanelli and Anthony Hastir for useful comments and discussions. R.M. is supported by a FRIA-FNRS PhD fellowship, grant FC 33443, funded by the Walloon region. \textcolor{black}{We also thank an anonymous referee for pointing out the reference~\cite{toth} that inspired us to prove the claim presented in the Remark~\ref{re:kinlin} and proved in the Appendix~\ref{sec:Applinkin}.}

%\disclaimer{Insert disclaimer text here.}

\appendix

\section{The Routh-Hurwitz criterion}
\label{sec:AppRHcrit}
\noindent
The Routh-Hurwitz criterion~\cite{Routh1877,Hurwitz1895,Barnett1983} is a well known tool of dynamical systems and control theory, allowing to prove the linear (in)stability of an equilibrium for a time invariant system. Indeed, the latter relies on the spectral properties of the Jacobian matrix evaluated at the \textcolor{black}{sought} equilibrium, which ultimately accounts to determine the location in the complex plane of the roots of a suitable polynomial.

We hereby present the method using a fourth order polynomial, but it applies to any given order ones. Let thus $p(\lambda)=a\lambda^4+b\lambda^3+c\lambda^2+d\lambda+e$ be a polynomial with real coefficients and assume $a>0$; the Routh-Hurwitz criterion can be stated using the Hurwitz matrix associated to $p(\lambda)$ and then compute its leading principal minors or building the Routh-Hurwitz  table and check the signs of the first column.

More precisely, a necessary condition\textcolor{black}{, also known in the literature as Stodola criterion~\cite{Bissell},} for the roots of $p(\lambda)$ to have negative real part is that all the coefficients are positive:
\begin{equation}
\label{eq:RHnec}
a>0 \, ,b>0\, ,c>0\,,d>0 \text{ and }e>0\, ,
\end{equation}
while a sufficient condition is
\begin{equation}
\label{eq:RHsuff}
a>0 \, ,b>0\, ,bc-da>0\,,d(bc-da)-eb^2>0 \text{ and }e>0\, .
\end{equation}

\subsection{Application of the criterion to the stability of $p_1(\lambda)$}
\label{ssec:RHstabp0}

Let us now apply the Routh-Hurwitz criterion to determine the stability feature of the polynomial $p_1(\lambda)$ given by Eq.~\eqref{eq:Cattaneo7}, hereby rewritten
\begin{eqnarray*}
p_1(\lambda)&=&a\lambda^4+b\lambda^3+c_1\lambda^2+d_1\lambda+e_1\\
&=&\tau_u\tau_v\lambda^4+(\tau_u+\tau_v)\lambda^3+\left[ 1-\tau_u\partial_v g-\tau_v\partial_uf\right]\lambda^2+\left[-\mathrm{tr}(J_0)\right]\lambda+\det(J_0)\, .
\end{eqnarray*}

Being the coefficients of $a=\tau_u\tau_v$ and $b=\tau_u+\tau_v$ positive, the Routh-Hurwitz criterion rewrites:
\begin{equation*}
c_1>0\, ,d_1>0\, ,bc_1-d_1a>0\,,d_1(bc_1-d_1a)-b^2e_1>0 \text{ and }e_1>0\, .
\end{equation*}
Replacing the definition of the coefficients in the above equation, we straightforwardly obtain the five conditions~\eqref{cond11}--\eqref{cond15}.

\subsection{Application of the criterion to the stability of $p_\alpha(\lambda)$}
\label{ssec:RHstabpalpha}

Let us now study the instability character of $p_\alpha(\lambda)$, for some $\alpha >1$ under the assumption of stability for $p_1(\lambda)$. We once again rely on the Routh-Hurwitz criterion. Let us thus rewrite
\begin{equation*}
p_\alpha(\lambda)=a\lambda^4+b\lambda^3+c_\alpha\lambda^2+d_\alpha\lambda+e_\alpha\, ,
\end{equation*}
where again $a=\tau_u \tau_v$, $b=\tau_u+\tau_v$ and
\begin{eqnarray*}
c_\alpha &=& 1-\tau_u\partial_v g-\tau_v\partial_uf -\Lambda^{(\alpha)}\left(\tau_uD_v+\tau_v D_u\right)=c_1-\Lambda^{(\alpha)}\left(\tau_uD_v+\tau_v D_u\right)\\ d_\alpha &=& -\mathrm{tr}(J_0)-\Lambda^{(\alpha)}(D_v+D_u)=d_1-\Lambda^{(\alpha)}(D_v+D_u)\\ e_\alpha &=& \det(J_0)+\left(D_v\partial_u f+D_u\partial_v g\right)\Lambda^{(\alpha)}+D_uD_v\left(\Lambda^{(\alpha)}\right)^2\, ,
\end{eqnarray*}
where we emphasised the relation between the coefficients defined for $\alpha>1$ and those for $\alpha=1$.

As already observed, the coefficients $a$ and $b$ are positive. Moreover, because of the assumption on the stability of $p_1(\lambda)$, we also have $c_1>0$ and $d_1>0$. Finally, observing that $-\Lambda^{(\alpha)}\geq 0$ for all $\alpha$ we can conclude that
\begin{equation*}
c_\alpha>0 \text{ and }d_\alpha>0\quad \forall \alpha\, .
\end{equation*}

The Routh-Hurwitz criterion ensures that $p_1(\lambda)$ is unstable if at least one of the following conditions is met:
\begin{itemize}
\item[i)] $bc_\alpha-d_\alpha a<0$;
\item[ii)] $d_\alpha(bd_\alpha-d_\alpha a)-b^2 e_\alpha<0$;
\item[iii)] $e_\alpha<0$.
\end{itemize}
Let us first show that condition i) is never met under the assumption of stability of $p_1(\lambda)$. From the definitions of the coefficients $a$, $b$, $c_\alpha$ and $d_\alpha$ we obtain
\begin{eqnarray*}
bc_\alpha-d_\alpha a&=& (\tau_u+\tau_v)\left[c_1-\Lambda^{(\alpha)}\left(\tau_uD_v+\tau_v D_u\right)\right]- \tau_u\tau_v \left[d_1-\Lambda^{(\alpha)}(D_v+D_u)\right]\\
&=&(\tau_u+\tau_v)c_1-\tau_u\tau_v d_1-\Lambda^{(\alpha)}\left[(\tau_u+\tau_v)\left(\tau_uD_v+\tau_v D_u\right)-\tau_u\tau_v(D_v+D_u)\right]\\
&=&bc_1-d_1 a-\Lambda^{(\alpha)}\left(\tau_u^2D_v+\tau_v^2 D_u\right)\, .
\end{eqnarray*}
We can now conclude that $bc_\alpha-d_\alpha a>0$; indeed because of the stability of $p_1(\lambda)$, $bc_1-d_1 a>0$, and being $-\Lambda^{(\alpha)}>0$ for all $\alpha>1$, the claim easily follows.

Let us now consider condition ii) and look for the existence of $\alpha>1$ such that 
\begin{equation*}
 d_\alpha(bc_\alpha-d_\alpha a)-b^2 e_\alpha<0\, .
\end{equation*}
We firstly rewrite this equation by using the definition of the involved coefficients
\begin{eqnarray*}
&\left(d_1-\Lambda^{(\alpha)}(D_v+D_u)\right)\left[(\tau_u+\tau_v)\left(c_1-\Lambda^{(\alpha)}\left(\tau_uD_v+\tau_v D_u\right)\right)-\left(d_1-\Lambda^{(\alpha)}(D_v+D_u)\right) \tau_u\tau_v)\right]+&\\
& -(\tau_u+\tau_v)^2\left[ \det(J_0)+\left(D_v\partial_u f+D_u\partial_v g\right)\Lambda^{(\alpha)}+D_uD_v\left(\Lambda^{(\alpha)}\right)^2\right]<0\,& ,
\end{eqnarray*}
and then we reorganise the terms in the latter, to write it as a second order polynomial in the variable $\Lambda^{(\alpha)}$, hence:
\begin{equation*}
A\left(\Lambda^{(\alpha)}\right)^2+B\Lambda^{(\alpha)}+C<0\, ,
\end{equation*}
where (after some algebraic manipulation):
\begin{eqnarray*}
 A&=&(\tau_uD_v-\tau_vD_u)^2\\
 B&=&-(\tau_u+\tau_v)(D_u+D_v)(1-\tau_u\partial_v g-\tau_v\partial_u f)+(\tau_u+\tau_v)\mathrm{tr}(J_0)(\tau_u D_v+\tau_v D_u)\\&-&2\mathrm{tr}(J_0)(D_u+D_v)\tau_u\tau_v-(\tau_u+\tau_v)^2(D_v\partial_u f+D_u\partial_v g)\\
 C&=&d_1(bc_1-ad_1)-b^2e_1\, .
\end{eqnarray*}
The coefficient $A$ is positive, as well as the coefficient $C$, under the assumption of stability for $p_1(\lambda)$. Then the second order polynomial in $\Lambda^{(\alpha)}$ can exhibit negative values if and only if
\begin{equation*}
 B>0 \text{ and }B^2-4AC>0\, ,
\end{equation*}
that are exactly the conditions~\eqref{eq:inst21} and~\eqref{eq:inst22}. Finally, an eigenvalue $\Lambda^{(\bar{\alpha})}$, $\bar{\alpha}>1$, must exist such that 
\begin{equation*}
x_1<\Lambda^{(\bar{\alpha})}<x_2\, ,
\end{equation*}
where $x_1$ and $x_2$ are the two real and negative roots of second order polynomial in $\Lambda^{(\alpha)}$.

Let us finally consider condition iii) and look for the existence of $\alpha>1$ such that 
\begin{equation*}
e_\alpha = \det(J_0)+\left(D_v\partial_u f+D_u\partial_v g\right)\Lambda^{(\alpha)}+D_uD_v\left(\Lambda^{(\alpha)}\right)^2<0\, .
\end{equation*}
This is a second order polynomial in the variable $\Lambda^{(\alpha)}$ whose leading coefficient, $D_uD_v$, is positive as well as the constant term, $\det(J_0)$, because of the stability of $p_1(\lambda)$. The polynomial can thus assume negative values if and only if
\begin{eqnarray*}
D_v\partial_u f+D_u\partial_v g&>&0\\
\left(D_v\partial_u f+D_u\partial_v g\right)^2-4D_uD_v\det(J_0)&>&0\, .
\end{eqnarray*}
Namely, the conditions~\eqref{eq:inst11} and~\eqref{eq:inst12}. Let us observe that the latter do not depend on $\tau_u$ and $\tau_v$ and indeed they are the classical conditions required for the Turing instability to arise~\cite{NM2010}: an eigenvalue $\Lambda^{(\bar{\alpha})}$, $\bar{\alpha}>1$, must exist such that 
\begin{equation*}
\eta_1<\Lambda^{(\bar{\alpha})}<\eta_2\, ,
\end{equation*}
where $\eta_1$ and $\eta_2$ are the two real and negative roots of $e_\alpha=0$.

\section{Non-relativistic limit of inertia-driven instability}
\label{sec:Appnonrellim}

In the main text we have proved that a Turing instability sets up driven by the inertial times $\tau_u$ and $\tau_v$ if any couple of conditions Eqs.~\eqref{eq:inst21} and~\eqref{eq:inst22} or Eqs.~\eqref{eq:inst11} and~\eqref{eq:inst12} are satisfied. Let us observe that the latter do not depend on the inertial times and thus they can be satisfied for a suitable choice of the model parameters, also for $\tau_u=\tau_v=0$. The same could not hold true for the former one, explicitly dependent on the inertial times. The aim of this section is thus to study the {\em non-relativistic limit} of the inertia-driven instability.

For a sake of definitiveness let us assume $\tau_v = \theta \tau_u$ for some $\theta>0$, that is the inertial times approach zero with the same rate. The characteristic polynomial given by~\eqref{eq:Cattaneo7} can thus be rewritten as
\begin{equation}
\label{eq:Cattaneo7b}
p_\alpha(\lambda)=\theta \tau_u^2\lambda^4+\tau_u(1+\theta)\lambda^3+(1-\tau_u \hat{c}_\alpha)\lambda^2+d_\alpha\lambda+e_\alpha\, ,
\end{equation}
where we have used~\eqref{eq:AB} to rewrite the coefficients of $\lambda^4$ and $\lambda^3$, and we have defined $\hat{c}_\alpha = \partial_v g+\theta \partial_uf +\Lambda^{(\alpha)}\left(D_v+\theta D_u\right)$ (see~\eqref{eq:C}). Let us observe that $d_\alpha$ and $e_\alpha$ do not depend on the inertial times (see~\eqref{eq:D} and~\eqref{eq:E}).

In the limit $\tau_u \rightarrow 0$ the latter results to be a singular polynomial, indeed its degree jumps from $4$ once $\tau_u>0$ to $2$ for $\tau_u=0$. Mathematically this means that two of the four roots of $p(\lambda)$ should diverge to infinity, by determining which ones and the followed path will allow to conclude about the non-relativistic limit of the inertia-driven instability.

Let us start by looking at the roots that remain in a bounded domain. To do this let us set~\footnote{To lighten the notation, we will not explicitly write the dependence on $\Lambda^{(\alpha)}$, our results will thus hold for all $\alpha=1,\dots, n$.} $\lambda = \lambda_0+\tau_u \lambda_1 +\dots$, impose $p(\lambda)=0$ and by reordering the involved terms (see Eq.~\eqref{eq:Cattaneo7b}) according to the powers of $\tau_u$, we eventually get
\begin{equation}
\label{eq:rel1}
0=p_\alpha(\lambda)=\lambda_0^2+d_\alpha\lambda_0+e_\alpha+\tau_u\left[(1+\theta)\lambda_0^3+2\lambda_0\lambda_1-\hat{c}_\alpha\lambda_0+d\lambda_1\right]  + \mathcal{O}(\tau_u^2)\, .
\end{equation}
We can thus conclude that $\lambda_0$ is a solution of the second degree equation
\begin{equation}
\label{eq:rel2}
\lambda_0^2+d_\alpha\lambda_0+e_\alpha=0\, ,
\end{equation}
while $\lambda_1$ is obtained by solving
\begin{equation}
\label{eq:rel3}
(1+\theta)\lambda_0^3+2\lambda_0\lambda_1-\hat{c}_\alpha\lambda_0+d\lambda_1=0\, .
\end{equation}
In conclusion we get for the two roots:
\begin{equation}
\label{eq:rel3b}
\lambda_{\pm} = \lambda_{0,\pm}\pm\tau_u \frac{\hat{c}_\alpha-(1+\theta)\lambda_{0,\pm}^2}{2\lambda_{0,\pm}+d_\alpha}+\mathcal{O}(\tau_u^2)\, ,
\end{equation}
where we denoted by $\lambda_{0,\pm}$ the two roots of~\eqref{eq:rel2}.
Let us observe that the latter is the same second order equation one will obtain in the classical Turing framework; we have thus shown that in the non-relativistic limit two roots of the fourth degree characteristic polynomial $p(\lambda)$ converge to the roots of the second order polynomial one should deal with in the classical Turing case.

Let us now study the remaining two roots and determine their path toward infinity. As already stated the characteristic polynomial is singular, one should thus resort to the {\em singular perturbation theory}~\cite{BenderOrszag}. Let us set $\lambda = \omega/\tau_u$ and evaluate $p_\alpha(\lambda)$ on $\lambda=\omega/\tau_u$, then we get
\begin{eqnarray}
\label{eq:rel4}
p_\alpha(\omega/\tau_u)&=&\frac{\theta}{\tau_u^2}\omega^4+\frac{1+\theta}{\tau_u^2}\omega^3+\frac{1-\tau_u\hat{c}_\alpha}{\tau_u^2}\omega^2+\frac{d_\alpha}{\tau_u}\omega+e_\alpha\notag\\&=&\frac{1}{\tau_u^2}\left[{\theta}\omega^4+(1+\theta)\omega^3+(1-\tau_u\hat{c}_\alpha)\omega^2+{d_\alpha}{\tau_u}\omega+e_\alpha \tau_u^2\right]=\frac{1}{\tau_u^2} q_\alpha(\omega)\, ,
\end{eqnarray}
where the fourth degree polynomial $q_\alpha(\omega)$ has been defined by the last equality. Let us observe that $p_\alpha(\lambda)$ vanishes if and only if $q_\alpha(\omega)$ does.

Let us now assume~\footnote{Let us stress once again that to lighten the notation we did not explicitly write the dependence on $\alpha$.} $\omega = \omega_0+\omega_1\tau_u+\mathcal{O}(\tau_u^2)$, with $\omega_0\neq 0$. By inserting the former into $q_\alpha(\omega)$ and by reordering the terms according to the powers of $\tau_u$, we get
\begin{equation}
\label{eq:rel5}
0=q_\alpha(\omega)={\theta}\omega_0^4+(1+\theta)\omega_0^3+\omega_0^2+ \tau_u \left[4\theta \omega_0^3\omega_1+3(1+\theta)\omega_0^2\omega_1+2\omega_0\omega_1-\hat{c}_\alpha\omega_0^2+d_\alpha \omega_0\right] + \mathcal{O}(\tau_u^2)\, .
\end{equation}
Hence, $\omega_0\neq 0$ solves the second degree equation
\begin{equation}
\label{eq:rel6}
\theta\omega_0^2+(1+\theta)\omega_0+1=0\, ,
\end{equation}
while $\omega_1$ is obtained by solving
\begin{equation}
\label{eq:rel7}
\omega_1\left[4\theta \omega_0^2+3(1+\theta)\omega_0+2\right]=\hat{c}_\alpha \omega_0-d_\alpha \, .
\end{equation}
In conclusion if $\theta >1$ we obtain
\begin{eqnarray}
\label{eq:rel8a}
\lambda_{+} &=& \frac{1}{\tau_u} \left[-\frac{1}{\theta}+\tau_u \frac{\hat{c}_\alpha+d_\alpha\theta}{\theta-1}+\mathcal{O}(\tau_u^2)\right]\\
\label{eq:rel8b}
\lambda_{-} &=& \frac{1}{\tau_u} \left[-1+\tau_u \frac{\hat{c}_\alpha+d_\alpha\theta}{1-\theta}+\mathcal{O}(\tau_u^2)\right]\, ,
\end{eqnarray}
while if $\theta <1$ we obtain
\begin{eqnarray}
\label{eq:rel9a}
\lambda_{+} &=& \frac{1}{\tau_u} \left[-1+\tau_u \frac{\hat{c}_\alpha+d_\alpha\theta}{1-\theta}+\mathcal{O}(\tau_u^2)\right]\\
\label{eq:rel9b}
\lambda_{-} &=& \frac{1}{\tau_u} \left[-\frac{1}{\theta}+\tau_u \frac{\hat{c}_\alpha+d_\alpha\theta}{\theta-1}+\mathcal{O}(\tau_u^2)\right]\, .
\end{eqnarray}
In both case we have that $\Re \lambda_\pm \rightarrow -\infty$ in the limit $\tau_u\rightarrow 0$ and thus these two roots cannot modify the (un)stable character of the homogeneous equilibrium.

In conclusion if $\tau_u$ and $\tau_v$ are positive and sufficiently small, then the onset of Turing instability is ruled out by the roots~\eqref{eq:rel3b}, i.e., those associated to the ones arising in the classical setting. Stated differently, if for $\tau_u>0$ and $\tau_v>0$, the instability can be initiated by conditions Eqs.~\eqref{eq:inst21} and~\eqref{eq:inst22}, then by decreasing the inertial times the patterns fade out and disappear before reaching the limit and thus the transition is not abrupt.

In Fig.~\ref{fig:tau1tau2} we report numerical results to complement the analytical findings  described above. We selected two generic sets of \textcolor{black}{parameter} values $\gamma = 4.0$, $\beta = 0.6$, $\mu = 1.0$, $D_u = 2.2$ and $D_v = 0.2$ (left panel) and $\gamma = 4.0$, $\beta = 0.7$, $\mu = 1.15$, $D_u = 0.2$ and $D_v = 2.2$ (right panel), and we study the emergence of an inertia-driven instability as a function of $\tau_u$ and $\tau_v$, a black dot corresponds to the presence of the instability while a white one to its absence. We can observe that the first set of parameters does not allow the onset of the instability for small enough values of the inertial times, indeed the black region, bounded below by condition~\eqref{cond15} and above by condition~\eqref{eq:inst22}, does not intersect the axes $\tau_u=0$ or $\tau_v=0$. On the other hand the second set of parameters allows the existence of patterns for $\tau_0=0$ or $\tau_v=0$, the black region (bounded above by condition~\eqref{eq:inst22}) reaches the axes. This means that in this case also classical Turing patterns are allowed.
\begin{figure}[ht]
\centering
\includegraphics[scale=0.25]{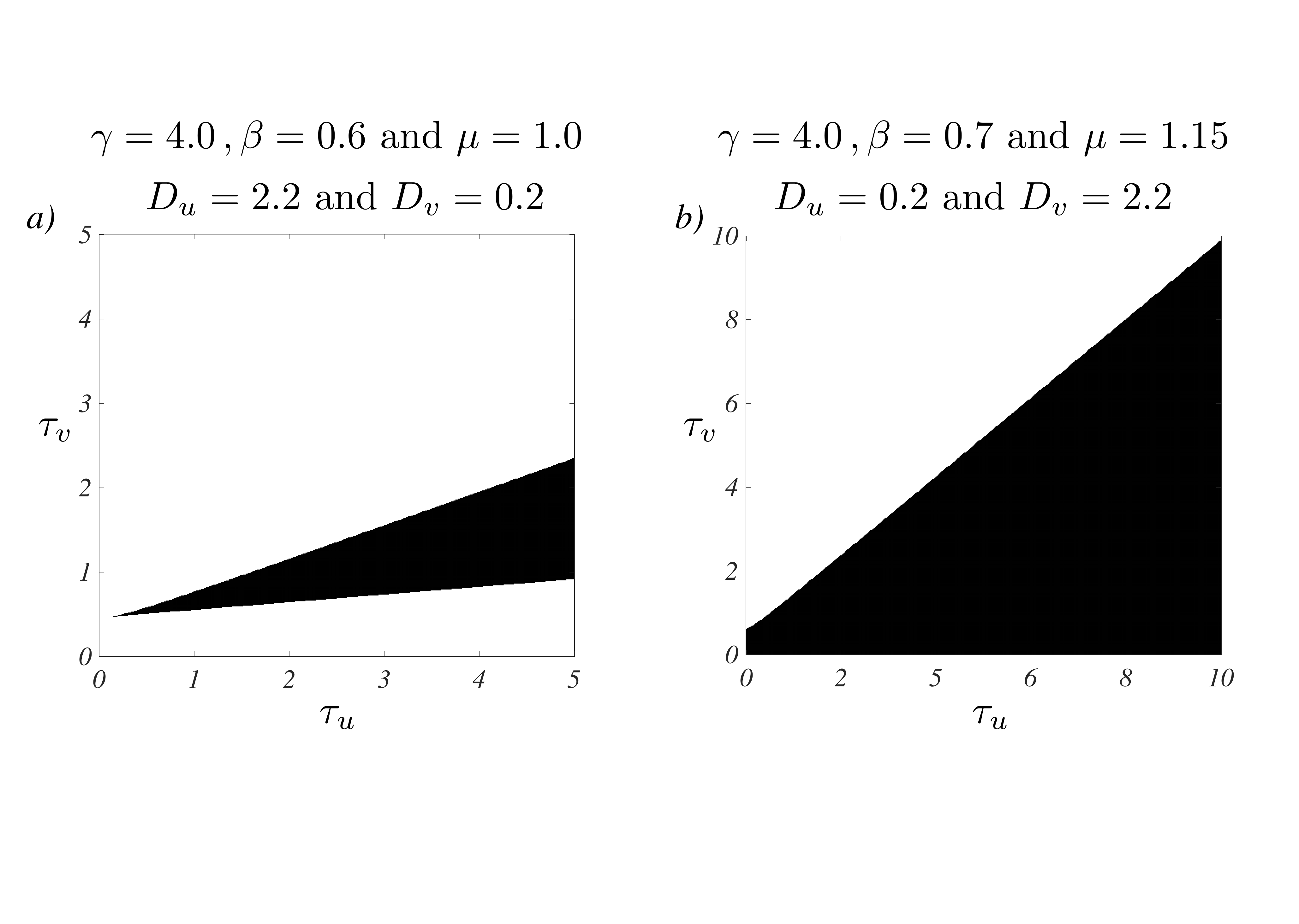}
\vspace{-1.5cm}
\caption{\textbf{Non-relativistic limit of inertia-driven instability for the {\em FHN} model.} For a fixed value of the model parameters $\gamma$, $\mu$, $\beta$, $D_u$ and $D_v$, we study the onset of inertia-driven instability (black regions) close to the homogeneous equilibrium $(u_i,v_i)=(0,0)$, $i=1,\dots,n$, as a function of $\tau_u$ and $\tau_v$. Panel a) corresponds to a setting where classical Turing instability cannot emerge, one can observe that the black region (bounded below by Eq.~\eqref{eq:inst22} and above by Eq.~\eqref{cond15}) does not touch the boundary $\tau_u=0$ or $\tau_v=0$. Panel b) refers to a choice of the parameters for which the classical Turing instability can emerge, indeed the black region (bounded above by Eq.~\eqref{cond15}) reaches the boundary $\tau_u=0$ or $\tau_v=0$.}
\label{fig:tau1tau2}
\end{figure}

\textcolor{black}{
\section{Existence of Turing instability in linear kinetic systems}
\label{sec:Applinkin}
Let us consider a linear kinetic system involving $m$ chemical species interacting each other ``inside'' the network nodes and allowed to diffuse across the available links under the constraint of finite propagation. Assuming the network to be composed by $n$ node and to denote the concentration of the $m$ species inside the $i$--th node by $\vec{x}^{(i)}=(x_1^{(i)},\dots,x^{(i)}_m)^\top\in\mathbb{R}_+^m$, then the main system~\eqref{eq:TRelnet} rewrites as
\begin{equation}
\label{eq:TRelnetKlin}
\frac{d{x}^{(i)}_j}{dt}+\tau_j \frac{d^2{x}^{(i)}_j}{dt^2}=\sum_{l=1}^m a_{jl}{x}^{(i)}_l+D_j\sum_{k=1}^{n}L_{ik} x^{(k)}_j  \quad\forall i=1,\dots,n \text{ and }\forall j=1,\dots,m \, ,
\end{equation}
where $D_j>0$(resp. $\tau_j>0$ ) is the diffusion coefficient (resp. the inertial time) for species $j=1,\dots,m$; $a_{jl}$ the linear kinetic representing the action of species $l$ on the $j$--th one and $L_{ik}$ the Laplace matrix encoding the network links.}

\textcolor{black}{To handle this equation, we decompose $\vec{x}^{(i)}$ on the Laplace eigenbasis, namely $x^{(i)}_j=\sum_\alpha \hat{x}_j^\alpha \phi_i^{(\alpha)}$, we insert the latter into Eq.~\eqref{eq:TRelnetKlin} and making use of the orthonormality of the eigenvectors we eventually obtain 
\begin{equation}
\label{eq:TRelnetKlinproj}
\frac{d\hat{x}^{\alpha}_j}{dt}+\tau_j \frac{d^2\hat{x}^{\alpha}_j}{dt^2}=\sum_{l=1}^m a_{jl}\hat{x}^{\alpha}_l+D_j\Lambda^{(\alpha)} \hat{x}^{\alpha}_j  \quad\forall \alpha=1,\dots,n \text{ and }\forall j=1,\dots,m \, .
\end{equation}}

\textcolor{black}{
The previous equation can be cast in a compact form by introducing $\vec{\hat{x}}^{\alpha}=(\hat{x}^{\alpha}_1,\dots,\hat{x}^{\alpha}_m)^\top$, the matrix of the diffusive coefficients $\mathbf{D}=\mathrm{diag}(D_1,\dots,D_m)$, the inertial times matrix $\mathbf{T}=\mathrm{diag}(\tau_1,\dots,\tau_m)$, and the linear kinetic matrix $\mathbf{a}$:
%\begin{equation}
%\label{eq:TRelnetKlinproja}
%\mathbf{T}\frac{d^2\vec{\hat{x}}^{\alpha}}{dt^2}+\frac{d\vec{\hat{x}}^{\alpha}}{dt}-\left(\mathbf{a}+\Lambda^{(\alpha)}\mathbf{D}\right)\vec{\hat{x}}^{\alpha}\quad\forall \alpha=1,\dots,n\, ,
%\end{equation}
%and left multiply the previous equation by $\mathbf{T}^{-1}$.
\begin{equation}
\label{eq:TRelnetKlinprojaa}
\frac{d^2\vec{\hat{x}}^{\alpha}}{dt^2}+\mathbf{T}^{-1}\frac{d\vec{\hat{x}}^{\alpha}}{dt}-\left(\mathbf{T}^{-1}\mathbf{a}+\Lambda^{(\alpha)}\mathbf{T}^{-1}\mathbf{D}\right)\vec{\hat{x}}^{\alpha}=0\quad\forall \alpha=1,\dots,n\, .
\end{equation}
Let us assume the matrix $\mathbf{A}:=\mathbf{T}^{-1}\mathbf{a}$ to be linearisable, i.e., there exist an invertible matrix $\mathbf{P}$ such that $\mathbf{P}^{-1}\mathbf{A}\mathbf{P}=\mathrm{diag}(\kappa_1,\dots,\kappa_m)$. Hence, by defining $\vec{y}=\mathbf{P}^{-1}\vec{x}$, and recalling that $\Lambda^{(1)}=0$, we can obtain from~\eqref{eq:TRelnetKlinprojaa}
\begin{equation}
\label{eq:TRelnetKlinprojb}
\frac{d^2\vec{\hat{y}}}{dt^2}+\mathbf{P}^{-1}\mathbf{T}^{-1}\mathbf{P}\frac{d\vec{\hat{y}}}{dt}-\mathrm{diag}(\kappa_1,\dots,\kappa_m)\vec{\hat{y}}=0\, .
\end{equation}
To make some analytical progress, let us assume all the inertial times to be equal each other, i.e., $\tau_j=\tau$ for all $j=1,\dots,m$. We have thus to solve $m$ second order linear ODE with constant coefficients, depending each one on $\kappa_j$
\begin{equation}
\label{eq:TRelnetKlinprojbb}
\frac{d^2{\hat{y}_j}}{dt^2}+\frac{1}{\tau}\frac{d{\hat{y}_j}}{dt}-\kappa_j{\hat{y}_j}=0\quad\forall j=1\dots,m\, .
\end{equation}
The associated characteristic polynomial is thus $\lambda^2+\frac{1}{\tau}\lambda-\kappa_j=0$, with roots  $\lambda = -\frac{1}{2\tau}\pm \frac{1}{2\tau} \sqrt{1+4\tau^2\kappa_j}$. One can prove that if $\Re \kappa_j<0$ and $\tau |\Im \kappa_j| < \sqrt{-\Re \kappa_j}$ then $\Re\lambda<0$.
}

\textcolor{black}{Assuming the linear kinetic does not admit cross inhibition terms, namely $a_{jl}\geq 0$ for all $j\neq l$, then we can conclude that the matrix $\mathbf{a}$ is non-negative and this implies that the matrix $\mathbf{A}=\mathbf{T}^{-1}\mathbf{a}$ is also essentially non-negative, being $\mathbf{T}$ a diagonal matrix with positive entries on the diagonal. Let us define $\mathbf{C}^{(\alpha)}=-\Lambda^{(\alpha)}\mathbf{T}^{-1}\mathbf{D}$. Then the latter is a diagonal matrix with non-negative elements on the diagonal, being $\Lambda^{(\alpha)}\leq 0$, $D_j>0$ and $\tau>0$. We can thus conclude invoking the following result proved in~\cite{toth}: if $\mathbf{A}$ is stable, i.e., its spectral abscissa in negative, then also $\mathbf{A}-\mathbf{C}$ is stable for all diagonal matrix $\mathbf{C}$ with non-negative diagonal terms. Indeed assume $\mathbf{A}=\mathbf{T}^{-1}\mathbf{a}$ to be stable, i.e. $\max_j \Re \kappa_j<0$, and moreover assume to fix $\tau$ such that $\tau |\Im \kappa_j| < \sqrt{-\Re \kappa_j}$ holds true. Let us observe that if $\mathbf{T}^{-1}\mathbf{a}$ has a real spectrum the latter relation is always satisfied. This implies that the homogeneous solution for the aspatial system is also stable. The Turing instability emerges if one can find $\alpha$ such that Eq.~\eqref{eq:TRelnetKlinprojaa} admits an unstable solution.}

\textcolor{black}{Let $\rho^{(\alpha)}_j$, $j=1,\dots,m$, be the eigenvalues of $\mathbf{A}-\mathbf{C}^{(\alpha)}=\mathbf{T}^{-1}\mathbf{a}+\Lambda^{(\alpha)}\mathbf{T}^{-1}\mathbf{D}$. Then the above quoted result~\cite{toth} ensures that $\Re\rho^{(\alpha)}_j<0$ for all $j=1,\dots,m$ and $\alpha=1,\dots,n$. Because $\mathbf{T}^{-1}\mathbf{D}$ is a diagonal matrix, the diagonal elements of $\mathbf{A}-\mathbf{C}^{(\alpha)}$ are translated to the left, while the off-diagonal elements do not vary. Invoking the Gershgorin circle theorem~\cite{golubvanloan}, we can take the elements of $\mathbf{T}^{-1}\mathbf{D}$ sufficiently large, i.e., $D_j\gg 1$ or $\tau \ll 1$, such that $\tau |\Im \rho_j^{(\alpha)}|<\sqrt{-\Re\rho_j^{(\alpha)}}$ for all $j=1,\dots,m$, $\alpha=1,\dots,n$. We can thus conclude that the relation dispersion associated to Eq.~\eqref{eq:TRelnetKlinprojaa} is always negative and thus the Turing instability cannot develop. In conclusion linear kinetic systems without cross inhibition and equal (small) inertial times, or large diffusion coefficients, cannot exhibit Turing instability in the relativistic framework. The necessity of the latter assumptions remains an open question that we believe goes beyond the scope of this work.}

\bibliographystyle{RS}
\bibliography{bib_RTP}

\end{document}